\documentclass[]{meta}

\usepackage{tikz}
\usetikzlibrary{arrows.meta,shapes.geometric,fit,backgrounds}
\definecolor{codeblue}{HTML}{4A90D9}
\definecolor{inputgray}{HTML}{95A5A6}
\definecolor{processgreen}{HTML}{27AE60}
\definecolor{specorange}{HTML}{F39C12}
\definecolor{testpurple}{HTML}{9B59B6}
\definecolor{successteal}{HTML}{1ABC9C}
\definecolor{refinered}{HTML}{E74C3C}
\makeatletter
\@ifpackageloaded{natbib}{}{\usepackage[numbers]{natbib}}
\makeatother
\usepackage{amsmath,amssymb,amsfonts}
\usepackage{algorithmic}
\usepackage{graphicx}
\usepackage{textcomp}
\usepackage{xcolor}
\usepackage{booktabs}
\usepackage{pdflscape}
\usepackage{pifont}
\makeatletter
\@ifpackageloaded{hyperref}{}{\usepackage{hyperref}}
\makeatother
\usepackage{balance}
\usepackage{enumitem}
\setlist{noitemsep, topsep=0pt, leftmargin=*}

\definecolor{metainternalgray}{HTML}{B0B0B0}
\newcommand{\metainternal}[1]{}%

\newcommand{\miN}{72}                            %
\newcommand{\miNcandidates}{122}                 %
\newcommand{\miSolveScore}{5.06}                 %
\newcommand{\miVC}{86.8\%}                       %
\newcommand{\miGT}{66.2\%}                       %
\newcommand{\miFlex}{72.1\%}                     %
\newcommand{\miAllThree}{54.4\%}                 %
\newcommand{\miDiffChars}{41{,}638}              %
\newcommand{\miVCavg}{10.5}                      %
\newcommand{\miTestAvg}{1.5}                     %
\newcommand{\ssRounds}{3}
\newcommand{\ssTotal}{72}
\newcommand{\ssConsidered}{68}
\newcommand{\ssNull}{4}
\newcommand{\ssStrengthened}{31}
\newcommand{\ssImproved}{26}
\newcommand{\ssBaseline}{5.06}
\newcommand{\ssBest}{5.74}
\newcommand{\ssDelta}{+0.68}

\newcommand{\toolname}{\textsc{AfterVibe}}
\newcommand{\datasetname}{TechInternal}

\def\BibTeX{{\rm B\kern-.05em{\sc i\kern-.025em b}\kern-.08em
    T\kern-.1667em\lower.7ex\hbox{E}\kern-.125emX}}

\definecolor{srcgray}{gray}{0.55}
\newcommand{\srcnote}[1]{{\color{srcgray}\scriptsize\,(#1)}}
\definecolor{specyellow}{HTML}{FFF7CC}
\newcommand{\elip}{[.\kern-0.13em.\kern-0.13em.]}
\newcommand{\specbox}[1]{%
  \par\smallskip\noindent
  \setlength{\fboxsep}{6pt}%
  \fcolorbox{black}{specyellow}{%
    \begin{minipage}{\dimexpr\columnwidth-2\fboxsep-2\fboxrule\relax}%
    #1%
    \end{minipage}}%
  \par\smallskip}

\definecolor{humangray}{gray}{0.90}
\newcommand{\humanbox}[1]{%
  \par\smallskip\noindent
  \setlength{\fboxsep}{6pt}%
  \fcolorbox{black}{humangray}{%
    \begin{minipage}{\dimexpr\columnwidth-2\fboxsep-2\fboxrule\relax}%
    #1%
    \end{minipage}}%
  \par\smallskip}

\usepackage{listings}

\let\cite\citep

\begin{document}

\title{\toolname{}: What Remains When the Conversation Ends}

\author{Matteo Paltenghi}
\author{Satish Chandra}
\affiliation{Meta, USA}

\correspondence{Matteo Paltenghi (\email{mattepalte@live.it}), Satish Chandra (\email{schandra@acm.org})}

\abstract{We present \toolname{}, a framework that recovers natural-language specifications from a vibe coding session.
Given a code artifact and the conversation trajectory that produced it, \toolname{} uses an LLM to extract an
abstract natural-language specification capturing the developer's intent, and validates it through a regeneration test: a
second, blind AI agent re-implements the artifact from the spec alone, and the resulting code is graded against
the original through a multi-tier validation pipeline. Spec quality is thus measured by whether an agent can
regenerate passing code; if the verifiers deem the implementations equivalent the spec is considered strong,
otherwise it is iteratively refined. Evaluating \toolname{} on \miN{} real-world vibe-coded projects from a company's
internal coding sessions, we find that its recovered specs are abstract by design---capturing behavioral intent
without dictating implementation---yet strong.
Multiple independent regenerations achieve a high mean regeneration score of \miSolveScore{} out of 6.0
while remaining diverse in their details, confirming that the spec constrains \emph{what} without
over-prescribing \emph{how}.
Besides outperforming existing human-authored descriptions, the specs can be further strengthened iteratively to a score of \ssBest{}.
A practical implication is that specifications---not code---could
become the primary artifact for human review and the source of record at a time when
AI-generated code is outpacing customary code review.
}

\maketitle

\section{Introduction}
\label{sec:intro}

The advent of powerful
large language models (LLMs) and coding agents has enabled a new style of programming in which
developers describe what they want in natural language and let an AI
assistant generate the code. Andrej Karpathy coined
the term \emph{vibe coding}\footnote{Andrej Karpathy, \emph{There's a New Kind of Coding Emerging}, X/Twitter post, February 2025.} to describe this practice: ``you fully give in to
the vibes, embrace exponentials, and forget that the code even exists.'' What
was once a niche experiment has rapidly become mainstream---millions of
developers now use AI coding assistants daily, and entire startups ship
products built almost exclusively through conversational prompting.
In this work, we use ``vibe coding'' broadly to denote any code change
produced through conversational interaction with an AI coding
agent---including production-quality changes at scale that are ultimately
reviewed and landed as accepted pull requests.

A well-known problem with vibe coding is
\emph{no one fully understands the code}---not even the developer who prompted it
into existence. Traditional software engineering relies on the assumption that
someone---be it the original author or a reviewer---can read, reason about,
and vouch for the correctness of source code. Vibe coding undermines this
assumption at its root. The developer's intent lives ephemerally in a chat
transcript; the generated code is voluminous and often opaque; and the mapping
between the two is implicit at best.

This situation is especially problematic for code review, one of the most
important quality-assurance practices in modern software
engineering~\cite{bacchelliExpectationsOutcomesChallenges2013}. A reviewer confronted with ever-growing
volume of AI-generated code has no concise, authoritative document that states
what the code is supposed to do.
The result is that a growing class
of software that is effectively \emph{unreviewable}, and not just because
of the rate of code production.  A recent industry study~\cite{Dora2025}
finds code review not keeping up with AI-based code production.

We propose \toolname, a framework that bridges the gap between vibe
coding and traditional quality assurance. The name reflects its
purpose: it captures what remains \emph{after} a vibe coding session
ends---a durable natural language specification distilled from the ephemeral
conversation. \toolname\ takes as input a
code artifact and the conversation trajectory that produced it, and outputs a
structured specification---what we call a \emph{spec}. The spec captures the developer's intent in natural language; crucially,
we treat requirements as implicit---whatever is observable from the
session is a sufficient expression of the requirements, since the
developer's actions during the session serve as a proxy for their mental
model.

To validate the faithfulness of the extracted specification,
\toolname\ employs a new \emph{regeneration test}: a second,
independent AI agent attempts to rebuild the artifact from the spec alone.
The regenerated code is then compared against the original artifact through
a three-tier verification pipeline: (1)~flexible test execution using
tests extracted from the session, (2)~verification conditions (VCs) extracted
independently from the conversation trajectory, and (3)~ground-truth
alignment checking via structured LLM
reasoning~\cite{ugareAgenticCodeReasoning2026}.
If the verifier deems the implementations equivalent, the spec is
considered faithful; if not, the
process iterates.
The regeneration test recalls the classical principle of design diversity and
N-version programming~\cite{chen1978nversion, knightExperimentalEvaluationAssumption1986}:
independent re-implementations of the same intent expose ambiguity and
unstated assumptions in a specification.

Our results show that natural-language specs derived from vibe coding
sessions are essentially sufficient to reconstruct the original code, are
abstract, and can be obtained as a byproduct of the vibe coding process rather
than requiring separate authoring effort.
Recovering requirements after implementation is the long-standing aim of
requirements traceability~\cite{gotelAnalysisRequirementsTraceability1994};
\toolname\ is the first to pursue it directly from human-AI coding trajectories.

\noindent \textbf{Contributions}. This paper makes the following contributions:
\begin{itemize}[noitemsep, leftmargin=*, topsep=0pt]
    \item[$\star$] \textbf{Retrospective specification recovery.}
    We introduce \emph{retrospective specification recovery}, the first
    technique to reconstruct developer intent \emph{after} coding, directly
    from the \emph{agent trajectory} (the conversation and code diff) of an
    AI-assisted coding session.

    \item[$\star$] \textbf{The regeneration strength test.}
    We propose the \emph{regeneration test}, an oracle that deems a
    specification faithful when a \emph{blind} agent can regenerate
    functionally equivalent code from it alone, scored by a three-tier
    agent-as-a-judge verifier.

    \item[$\star$] \textbf{Evidence of real-world effectiveness.}
    We implement the technique in \toolname\ and evaluate it on \miN{}
    real-world, vibe-coded tasks from an industrial
    monorepo{} where recovered specifications reach a
    mean \textit{regeneration score} of \ssBest{}/6.0 after iterative strengthening.

    \item[$\star$] \textbf{Specs that are abstract yet sufficient.}
    We show that recovered specifications are abstract by design---independent
    regenerations succeed while differing substantially in implementation, so a
    spec constrains \emph{what} without over-prescribing \emph{how}.  The specs are
    also significantly more concise than the diff in size: average reduction of
    5.6x in chars.
\end{itemize}

\noindent \textbf{Vision}. Our vision is that \emph{specifications, not code, should be the primary
artifact of human authorship} in the age of AI-assisted development.
Code---whether written by a human or generated by an AI---is merely an
implementation of intent. By elevating specifications to first-class status,
we enable a new review paradigm in which humans inspect abstract, readable
statements of intent rather than voluminous, machine-generated code.
Beyond review, the specification can serve as the \emph{source of
record}---the canonical artifact from which an implementation can be
reconstructed, analogously to how executable code is compiled from
high-level languages.

This paper has shown the feasibility of recovering specifications from
vibe coding sessions automatically.  In future work, we plan to study
empirically whether and how well such specifications can get adopted
as the artifact to be reviewed and possibly a trusted source of record.

\section{Motivating Example}
\label{sec:motivating}

To ground the discussion, we walk through an example inspired by a real
internal coding session (anonymized). A developer
rewrites a \emph{resource-cleanup utility}: a component that, when a
logical tenant is torn down, deletes the tenant's leftover bookkeeping
records across several internal stores. The original implementation was
\emph{blind}---it issued a delete for every record type in every group
whether or not the record existed---wasting a rate-limited write budget.
The developer uses an AI coding assistant to refactor it into a
\emph{scan-then-delete} model and to fix a cluster of correctness and
robustness issues. The resulting change spans over 100{,}000
characters across many files.

\begin{figure}[t]
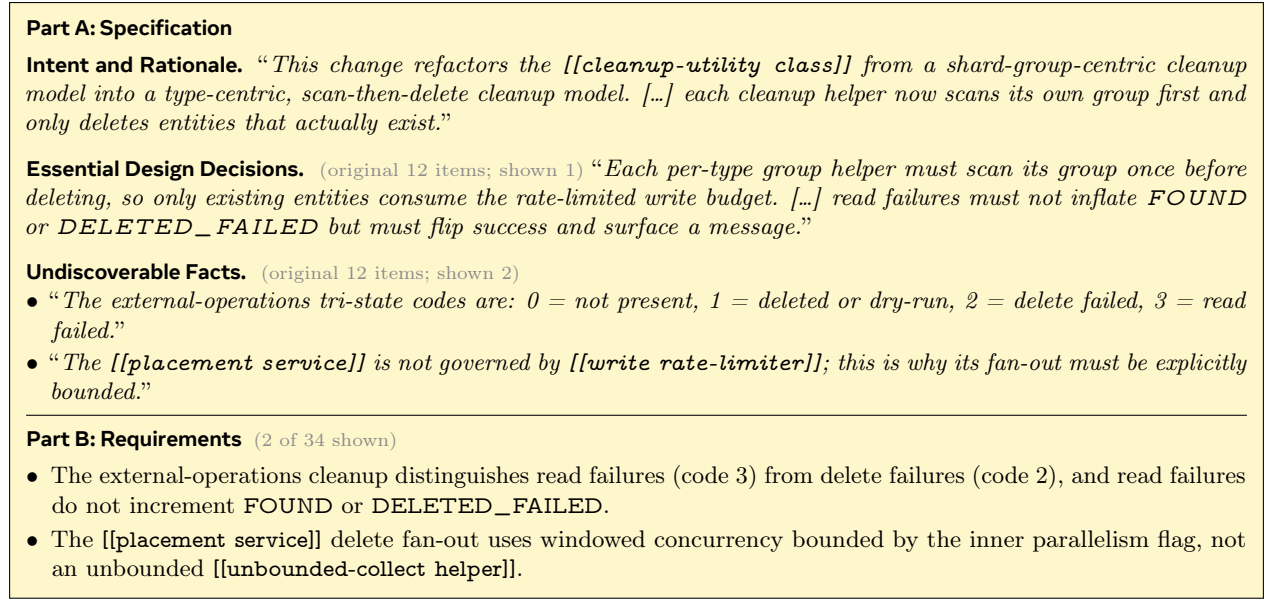

\specbox{\small
\textbf{Part~A: Specification}\par\smallskip
\textbf{Intent and Rationale.}
``\emph{This change refactors the \texttt{[[cleanup-utility
class]]} from a shard-group-centric cleanup model into a type-centric,
scan-then-delete cleanup model. \elip{} each cleanup helper now scans
its own group first and only deletes entities that actually exist.}''

\medskip
\textbf{Essential Design Decisions.}
\srcnote{original 12 items; shown 1}
``\emph{Each per-type group helper must scan its group once before
deleting, so only existing entities consume the rate-limited write
budget. \elip{} read failures must not inflate \texttt{FOUND} or
\texttt{DELETED\_FAILED} but must flip success and surface a message.}''

\medskip
\textbf{Undiscoverable Facts.}
\srcnote{original 12 items; shown 2}
\begin{itemize}\setlength{\itemsep}{1pt}
\item ``\emph{The external-operations tri-state codes are: 0 = not
present, 1 = deleted or dry-run, 2 = delete failed, 3 = read failed.}''
\item ``\emph{The \texttt{[[placement service]]} is not governed
by \texttt{[[write rate-limiter]]}; this is why its fan-out must
be explicitly bounded.}''
\end{itemize}

\medskip
\hrule
\medskip
\textbf{Part~B: Requirements}
\srcnote{2 of 34 shown}\par\smallskip
\begin{itemize}\setlength{\itemsep}{2pt}
\item The external-operations cleanup distinguishes read failures
(code 3) from delete failures (code 2), and read failures do not
increment \texttt{FOUND} or \texttt{DELETED\_FAILED}.
\item The \texttt{[[placement service]]} delete fan-out uses windowed
concurrency bounded by the inner parallelism flag, not an unbounded
\texttt{[[unbounded-collect helper]]}.
\end{itemize}
}
\caption{Full \toolname{} post-hoc spec (Part~A $+$ Part~B).}
\label{fig:spec}
\end{figure}

\begin{figure}[t]
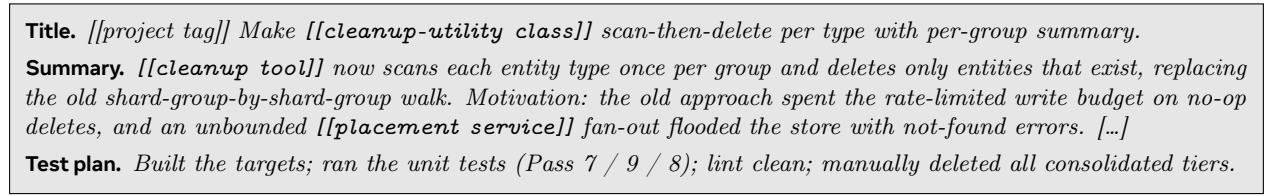

\humanbox{\small
\textbf{Title.} \emph{[[project tag]] Make \texttt{[[cleanup-utility
class]]} scan-then-delete per type with per-group summary.}

\smallskip
\textbf{Summary.} \emph{\texttt{[[cleanup tool]]} now scans each entity
type once per group and deletes only entities that exist, replacing the
old shard-group-by-shard-group walk. Motivation: the old approach spent
the rate-limited write budget on no-op deletes, and an unbounded
\texttt{[[placement service]]} fan-out flooded the store with not-found
errors. \elip{}}

\smallskip
\textbf{Test plan.} \emph{Built the targets; ran the unit tests
(Pass 7 / 9 / 8); lint clean; manually deleted all consolidated tiers.}
}
\caption{Full human-authored summary.}
\label{fig:human}
\end{figure}

\toolname{} takes the change and the conversation transcript as input
and produces the structured specification of Fig.~\ref{fig:spec}.\footnote{In the excerpts, \elip{} marks elided text and names in
double-bracketed \texttt{typewriter} font (e.g.\ \texttt{[[cleanup-utility
class]]}) are \emph{anonymized placeholders} for real internal identifiers.
Technical values (return codes, field keys, defaults) are non-identifying and
kept verbatim.} The full \toolname{} spec is
under 10{,}000 characters.
~
The full post-hoc specification (Fig.~\ref{fig:spec}) has two parts---\emph{Part~A:
Specification} (three prose sections) and \emph{Part~B: Requirements} (a
behavioral checklist of declarative statements)---which together we refer
to as the \textit{spec}, and which are what a blind agent regenerates
from.
~
For contrast, the developer's own review description---title, summary,
and test plan---used verbatim as a baseline ``spec''
(Fig.~\ref{fig:human}, shortened).

Both artifacts describe the same change, and the human summary is not
vague---it names the scan-then-delete idea, the typed result, and the
bounded fan-out. Yet the two rebuilds diverge sharply. The \toolname{}
rebuild passes both extracted tests ($2/2$), satisfies all twelve
verification conditions ($12/12$), and matches the reference on a holistic
review, for a perfect $6$ out of $6$; the summary's rebuild passes
neither test ($0/2$), meets only three of the twelve verification
conditions ($3/12$),
and fails the holistic review, scoring $0.5$. The reason is that a
changelog reports \emph{what happened} whereas a re-implementer needs the exact \emph{rules} to follow.
Two of the verification conditions---shown in Fig.~\ref{fig:vcs}---make this
concrete. For each, our spec states the exact rule (the distinct codes
returned for a failed \emph{read} versus a failed \emph{delete}, and that a
tenant is skipped only when \emph{both} group lists are empty), so the rebuilt
code satisfies the check. The summary omits both distinctions, so its rebuild
merges the two failure cases and stops as soon as \emph{either} list is empty,
failing both checks.
The same gap breaks the build itself: our spec names the exact build
dependency to switch, whereas the summary only says the data now comes
from a different source, so the rebuilt project files disagree with the
code and it never compiles---which is why neither of its tests can run.

\begin{figure}[t]
\specbox{\small
\textbf{Verification conditions.}
\begin{itemize}\setlength{\itemsep}{2pt}
\item Is a failed \emph{read} of a record counted separately from a
failed \emph{delete}, and never tallied as a delete failure?
\item Does the cleanup skip a tenant only when \emph{both} of its two
group lists are empty, still cleaning the other when only one is empty?
\end{itemize}
}
\caption{Verification conditions: checks the grader extracts
independently of the spec and runs on both the original and the rebuilt
code (2 of 12 shown).}
\label{fig:vcs}
\end{figure}

\section{Approach}
\label{sec:approach}

\begin{figure*}[t]
\centering
\includegraphics[width=\textwidth]{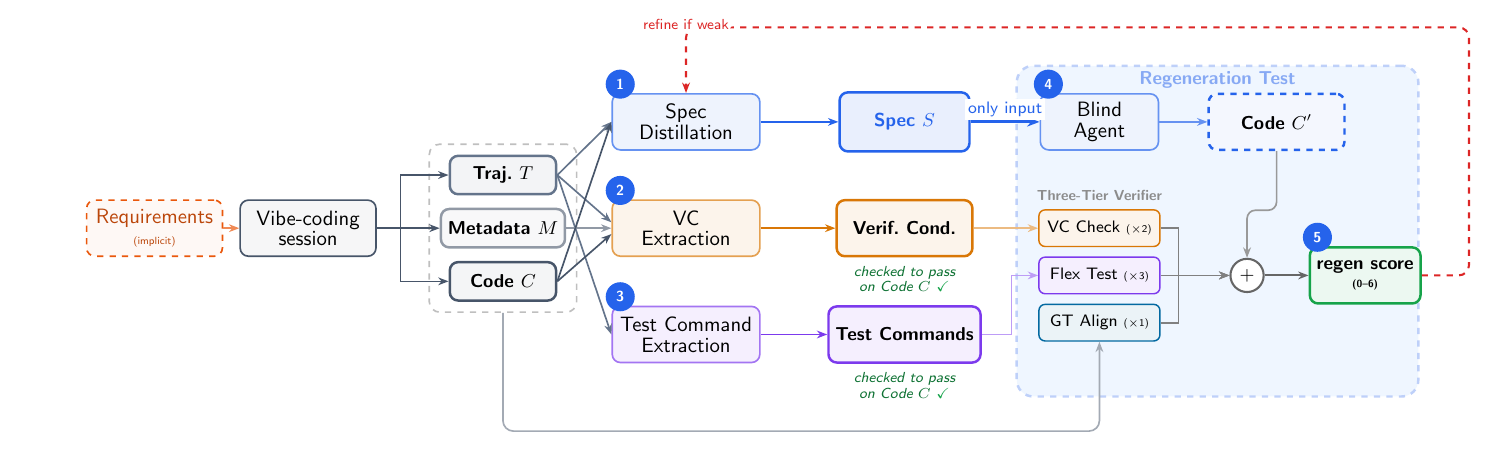}
\caption{Overview of the \toolname\ workflow.
From a vibe-coding session (trajectory~$T$, summary~$M$, code~$C$), \toolname\
\ding{192}~distills a specification~$S$;
\ding{193}~extracts verification conditions and
\ding{194}~test commands (both validated on~$C$ as oracles);
\ding{195}~regenerates code~$C'$ from~$S$ alone with a blind agent; and
\ding{196}~scores it with a three-tier verifier, iteratively refining~$S$ when
the score is weak.}
\label{fig:overview}
\end{figure*}

Figure~\ref{fig:overview} gives an overview of the \toolname\ workflow:
from a vibe-coding session it distills a specification, derives oracles
(verification conditions and test commands), regenerates code from the
specification alone, and scores the regeneration with a three-tier
verifier, iteratively refining the specification when the score is weak.
We detail each stage in the remainder of this section.

\subsection{Problem Formulation}
\label{sec:problem}

We define the retrospective specification recovery problem as follows.
Given a code change~$C$ (unified diff) produced by an AI coding assistant
against a repository at a base commit, and a conversation trajectory~$T$ between
the developer and the assistant (user turns, agent tool calls, and messages),
the goal is to produce a specification~$S$ that is both \emph{strong}---a
satisfactory implementation can be derived from~$S$ alone---and
\emph{abstract}---capturing behavioral intent without prescribing implementation.
These desiderata are in tension: implementation detail increases strength but
reduces abstractness, while aggressive abstraction risks omitting requirements.
A good specification is thus abstract where it can be and concrete only where it
must be, gated by what an agent can discover from the environment.

We do not require the developer to reify their requirements
explicitly; instead, we treat requirements as implicit---whatever is
observable from the session (the developer's prompts, corrections,
and accepted behaviors) is a sufficient expression of the
requirements.

We operationalize strength through the \emph{regeneration test}
(Section~\ref{sec:regen}): if a blind AI agent, starting from the same base
commit, can produce a code change $C'$ from $S$ alone such that a
three-tier verifier deems $C'$ equivalent to $C$ modulo the conversation $T$,
then $S$ is considered strong.
The verifier grounds its judgment in the task requirements expressed in $T$,
accepting any implementation that conforms to the stated intent regardless
of structural differences from $C$.

\paragraph{The environmental grounding hypothesis.}
\label{sec:grounding}
The abstractness desideratum rests on a hypothesis: a coding agent
can reconstruct substantial contextual information---build
configurations, naming conventions, module structure, API
contracts---by navigating the repository, rather than requiring the
specification to spell it out.
If this hypothesis holds, a spec needs to be concrete only for
\emph{undiscoverable} facts: decisions, thresholds, and domain
constraints that cannot be inferred from the codebase.
Everything else can be stated abstractly, because the agent's
environmental grounding fills the gap.

\subsection{Spec Distillation}
\label{sec:distillation}

The distillation stage transforms the raw inputs---the code change $C$ and
the conversation trajectory $T$---into an abstract natural-language
specification $S$.
We prompt the LLM with $C$ and $T$, along with an \emph{extraction prompt}
that instructs it to produce a structured specification in three prose
sections---\emph{Intent and Rationale}, \emph{Essential Design Decisions},
and \emph{Undiscoverable Facts}---together with a behavioral checklist
of requirements phrased as declarative statements (Figure~\ref{fig:master-prompt}).

The extraction prompt enforces abstractness by default: it instructs
the LLM to use ``no code, identifiers, paths, or line numbers unless
essential and undiscoverable,'' making concreteness the exception
rather than the rule.
The three sections operate at different abstraction levels:
\emph{Intent and Rationale} is maximally abstract (pure \emph{what}
and \emph{why}); \emph{Essential Design Decisions} sits at
mid-level (\emph{which approach}, but not how to code it);
\emph{Undiscoverable Facts} is necessarily concrete---names,
thresholds, and contracts that the agent cannot infer from the
environment.
Section~\ref{sec:motivating} shows a concrete example of a spec produced
by this prompt.

This design reflects a key trade-off.
A highly structured template (e.g., with separate sections for functional
requirements, data model, and error handling) would impose organization on
every change, but risks over-specifying simple patches and
under-representing concerns that do not fit the template.
By contrast, a free-form but intent-focused prompt lets the LLM adapt the
level of detail to the complexity of the change.

\begin{figure}[t]
\fbox{\begin{minipage}{\dimexpr\columnwidth-2\fboxsep-2\fboxrule}
\small\color{blue}
Given the following conversation and diff, create a specification that another coding agent can use to reimplement the change.

\medskip
\textbf{Part 1: Specification}\\
Write three short prose sections (no code, identifiers, paths, or line numbers unless essential and undiscoverable):
\begin{itemize}\setlength{\itemsep}{0pt}
\item \textbf{Intent and Rationale:} what the code changes achieve.
\item \textbf{Essential Design Decisions:} key behaviors/criteria that must be preserved (separate essential from incidental).
\item \textbf{Undiscoverable Facts:} external names, thresholds, contracts, and domain gotchas an implementer cannot infer from the codebase. Includes file paths or symbol names explicitly requested in the conversation---preserve verbatim.
\end{itemize}

\textbf{Part 2: Requirements}\\
List the behavioral requirements as concise declarative statements. Focus on outcomes, not implementation. Do NOT include instructions about tests.
\end{minipage}}
\caption{Extraction prompt for spec distillation (shortened for presentation).}
\label{fig:master-prompt}
\end{figure}

\subsection{Three-Tier Verification}
\label{sec:deep-verifier}

A key design question in \toolname\ is how to determine whether the
regenerated code $C'$ is functionally equivalent to the original artifact
$C$. We employ a three-tier verification pipeline that combines
complementary signals at increasing levels of semantic depth.
Each tier is implemented as an autonomous subagent with repository access,
placing the verifier collectively in the agent-as-a-judge
paradigm~\cite{zhugeAgentasaJudgeEvaluateAgents2025}---an extension of the
successful LLM-as-a-judge approach~\cite{zhengJudgingLLMasajudgeMTbench2023, guSurveyLLMasaJudge2025}.
To offset known judgment biases, no single tier is decisive: we combine
execution-grounded and reasoning-based signals.

\paragraph{Flexible test execution.}
The first tier runs tests derived from the conversation trajectory---commands
the developer executed during the vibe-coding session---against the
regenerated code~$C'$.
Because $C'$ may expose slightly different interfaces
than the original~$C$, \toolname\ employs a \emph{flexible test runner}
that adapts test harnesses to minor interface changes (e.g., renamed
entry points or reordered parameters) rather than failing on superficial
mismatches.
This flexibility avoids overly prescriptive testing while still
verifying that the regenerated code satisfies the behavioral expectations
embedded in the session.

\paragraph{Verification conditions.}
The second tier extracts \emph{verification conditions} (VCs)---testable
behavioral properties---from the conversation trajectory $T$, the code
change~$C$, and the human-authored summary~$M$ (the diff title,
description, and test plan the developer submits for code review).
VCs express what the code should do (e.g., ``the date-range picker filters all
three charts'') without prescribing how.
Crucially, VCs are not derived from the spec~$S$, which would make the
validation circular.
This independence ensures that VCs serve as an
unbiased oracle for assessing whether both $C$ and $C'$ satisfy the
developer's requirements.
This independence also lets the VCs act as a fixed reference point: as we
will see later, we iteratively refine the spec~$S$ to make it stronger,
but we hold the verification conditions constant throughout, so that every
candidate spec is judged against the same oracle.

\paragraph{Ground-truth alignment.}
Verification conditions check properties individually; ground-truth
alignment checks whether the regeneration makes sense as a whole.
Recent work on agentic code reasoning demonstrates that LLM agents can
perform meaningful semantic code analysis without executing
code~\cite{ugareAgenticCodeReasoning2026}.
~
The ground-truth alignment checker is an LLM agent that receives the
conversation trajectory~$T$, the original code change~$C$, the regenerated
code~$C'$, and full access to the repository at the base commit.
It determines whether $C'$ conforms to the intent expressed in~$T$ and
the behavioral expectations established by~$C$.
Equivalence is judged \emph{modulo the conversation}: because vibe-coded
changes are often broad and under-determined, there is no single correct
implementation.
Any regeneration that conforms to the intent and constraints expressed in the
conversation is considered valid, even if it differs substantially from the
original in structure, naming, or implementation strategy.

\paragraph{Combining tiers.}
The three complementary tiers each contribute a weighted score to a combined
\emph{regeneration score} (\text{regen\_score}, 0--6), defined as
$\text{regen\_score} = 3\,\mathit{flex} + 2\,\mathit{vc} + \mathit{align}$, where
$\mathit{flex} \in [0,1]$ is the fraction of flexible tests that pass,
$\mathit{vc} \in [0,1]$ is the fraction of verification conditions satisfied,
and $\mathit{align} \in \{0,1\}$ is the binary ground-truth alignment verdict.
The weights encode a deployment-informed priority: execution-grounded tests rank
highest~(3) as the strongest signal that the regeneration is not broken,
verification conditions moderate~(2) as independent checks needing no execution
infrastructure, and ground-truth alignment lowest~(1) to avoid over-rewarding
literal reproduction of the reference patch.
The weighting is a configurable design choice, and our findings are robust to it:
under equal weights, swapped flex/VC weights, and a ground-truth-heavy weighting,
the \toolname{}-vs-human comparison keeps the same sign and the per-task ranking
is preserved (Section~\ref{sec:rq6}).

\subsection{The Regeneration Test}
\label{sec:regen}

The regeneration test is the core validation mechanism of \toolname. A
fresh LLM instance---with no access to the original code $C$ or conversation
$T$---receives only the spec $S$ and is asked to implement the artifact
from scratch. The intuition is that if the spec is sufficiently strong, a
competent implementer (human or AI) should be able to produce a
functionally equivalent artifact.

Formally, we define the regeneration test as:
\begin{equation}
\text{RegenTest}(S, C, T) = \begin{cases}
    \text{PASS} & \text{if } V(C', C, T) > \tau \\
    \text{FAIL} & \text{otherwise}
\end{cases}
\label{eq:regen}
\end{equation}
where $V$ is the three-tier verifier (Section~\ref{sec:deep-verifier}),
$C'$ is the code produced by the blind regeneration agent,
$T$ is the conversation trajectory providing task context, and $\tau$ is a passing threshold. The verifier applies
test execution, verification-condition checking, and ground-truth
alignment, grounding its judgment in the task requirements expressed in $T$.

\subsection{Iterative Spec Refinement}
\label{sec:refinement}

When a task's regenerated code scores below a threshold
(regeneration score~$< \tau$, default $\tau = 6.0$), \toolname\ enters a
per-task \emph{spec strengthening} loop that refines the individual
specification rather than the global extraction prompt.
This feedback-driven refinement follows the iterative
generate--critique--improve paradigm shown to be effective for
LLM-based program repair, fuzzing, and general-purpose text
improvement~\cite{madaanSELFREFINEIterativeRefinement2023,xiaAutomatedProgramRepair2024,Fuzz4AllUniversalFuzzing}.
The strengthening operates at the granularity of individual
verification conditions: for each VC that the regenerated code
fails, a refinement LLM receives the original spec~$S$, the
ground-truth diff~$C$, the regenerated diff~$C'$, the failed
verification condition, and the grader's rejection reasoning.
It then analyzes why the regeneration diverged from the original
intent, identifies what was ambiguous or underspecified in~$S$, and
produces a complete rewritten spec~$S'$.
The refined spec is re-evaluated through the full pipeline: a fresh
agent regenerates code from~$S'$, and the three-tier verifier grades
the result.

\paragraph{Best-of-K selection.}
After $K$ strengthening rounds, \toolname\ keeps the \emph{best spec
per task}---the one that achieved the highest regeneration score across the
baseline and all rounds.
Only tasks scoring below~$\tau = 6.0$ enter the loop; once a task
reaches that threshold it exits early.
Because the selection takes the per-task
maximum, the aggregate score is monotonically non-decreasing by
construction: individual rounds may score below the baseline (each
is a single noisy rollout), but the best-of-K can only improve or
stay the same.
Note that strengthening quality is ultimately bounded by the
information in the conversation trajectory: a vague or underspecified
session yields a weak spec regardless of refinement effort.

\section{Evaluation}
\label{sec:evaluation}

We evaluate \toolname\ along six research questions:

\begin{itemize}
    \item \textbf{RQ1 -- Regeneration Fidelity:} Can a spec distilled from a
    vibe-coding session let a blind agent regenerate functionally equivalent code?
    \item \textbf{RQ2 -- Spec Strengthening:} Can per-task verifier feedback
    strengthen individual specifications and increase solve rate?
    \item \textbf{RQ3 -- Abstraction vs.\ Strength:} How does specification
    abstraction relate to regeneration fidelity, and what role does code-token leakage play?
    \item \textbf{RQ4 -- Regeneration Diversity:} How similar are passing
    regenerations from the same spec, relative to natural upper and lower bounds?
    \item \textbf{RQ5 -- Grader Validity:} Do the extracted verification conditions
    and test commands reliably separate correct code from wrong or degraded code?
    \item \textbf{RQ6 -- Comparison with Human-Authored Descriptions:} Does \toolname{}
    produce more effective specs than the commit descriptions developers already write?
\end{itemize}

\subsection{Dataset}
\label{sec:dataset}

\subsubsection{Company Internal Dataset (\datasetname{})}
\label{sec:dataset-cip}

Our primary evaluation uses real-world coding sessions from an internal
AI coding assistant that follows
a monorepo-based, incremental software development workflow.
{}
We refer to this dataset as \textbf{\datasetname{}}.

\paragraph{Source and curation.}
We sample multi-turn coding-assistant sessions
{}
from a 7-day window. Each session consists of a developer interacting with an
AI coding assistant to implement a code change in a large monorepo.
Sessions are selected by a curation funnel that filters for:
(i)~completed sessions with a landed diff,
(ii)~non-trivial code changes (excluding pure config or generated-file edits),
and (iii)~sessions with sufficient conversation context for spec extraction.
This funnel yields \miNcandidates{} candidate sessions.

\paragraph{Ground-truth validation.}
We validate each candidate through a ground-truth verification run:
the verification pipeline is executed against the \emph{original landed
code}---not a regeneration.
{}
Two graders run on the original code: the flexible test runner and the
verification conditions checker.
Ground-truth alignment is skipped because there is no regeneration to compare against.
Of the \miNcandidates{} candidates, 81 complete with both graders
producing a verdict (the remaining 41 are excluded due to
infrastructure failures: evaluation-sandbox crashes, grader-process
timeouts, or tool-server startup errors); the excluded tasks have
comparable verification-clue counts (median 11 vs.\ 11), test-manifest
counts (median 2 vs.\ 2), and similar or slightly higher session
durations (median 16\,min vs.\ 14\,min), consistent with
infrastructure retries and timeouts rather than a task-complexity bias.
Of the 81 fully graded candidates, \miN{} pass both graders
(88.9\% pass rate), forming the final evaluation set.

\paragraph{Failure analysis.}
The 9 ground-truth failures---sessions where the \emph{landed, CI-passing}
code fails at least one grader---stem from three causes.
Four are \textbf{test-command environment mismatches}: the LLM-extracted
test commands assume developer-machine dependencies (JavaScript runtimes,
native shared libraries, device interfaces, or build-graph targets)
absent in the evaluation sandbox.
Three are \textbf{test-extraction failures}: the LLM extracted zero test
commands or a nonexistent test target from the conversation transcript.
Two are \textbf{grading-rule edge cases}: the reference patch deletes the
only test file, so the grader marks it skipped and conservatively scores
the empty result as a failure.
None of these 9 represent code regressions---all \miNcandidates{} diffs
passed CI and landed successfully.
The failures reflect limitations of mining test oracles from conversation
context; environment-aware test-command extraction could recover several
of these tasks.

\paragraph{Characteristics.}
The \miN{} validated tasks span typical monorepo development activities:
bug fixes, feature additions, refactors, and infrastructure changes.
The average original diff is \miDiffChars{} characters.
Each task has on average \miVCavg{} verification
conditions extracted from the conversation trajectory and \miTestAvg{} test
commands derived from the session.
Both verification conditions and test commands are extracted by dedicated LLM-based pipelines.

\subsection{Experimental Setup}

All LLM components---spec distillation, verification condition and test command extraction, strengthening feedback, and the regeneration agent---use a single frontier large language model.
The regeneration agent runs in an agentic harness with code-editing tools, shell access, and automated graders that evaluate the result after each attempt.
{}
The distillation prompt follows the template described in
Section~\ref{sec:distillation}. The regeneration agent receives only the vibe
spec and a generic instruction to implement the described artifact.
Verification artifacts---verification conditions and test commands---are extracted
independently from the conversation trajectory using dedicated
LLM-based extraction pipelines.
{}
Each task is evaluated by the three active graders of the multi-tier
verification pipeline (Section~\ref{sec:deep-verifier}).
We report the \emph{regeneration score} (0--6 scale) as the primary metric.
~
For grader validity (Section~\ref{sec:rq5}), we also run spec-free negative controls to verify that the extracted verification artifacts have genuine discriminative power---that is, they pass on correct code but fail on wrong or degraded code.

\subsection{RQ1: Regeneration Fidelity}
\label{sec:rq1}

{}
{}

For each task, we distill a post-hoc specification from the original
conversation and code change~$C$.
A fresh agent then receives only this specification and attempts to
regenerate the target code change from the same base commit.
We compare the regenerated code $C'$ against $C$ using the three-tier
verification pipeline.

The mean regeneration score is \miSolveScore{}/6.0 (Table~\ref{tab:human-baseline}).
Scores are aggregated over the 68 of \miN{} tasks that produced a
complete grader verdict; the remaining four never reached grading and
are excluded from the scored aggregates.
Inspecting their trajectories shows these were all harness failures
rather than model or task failures: one event-serialization crash, one
wall-clock timeout, and two startup aborts, and three of the four
produced no patch at all.
The verification conditions grader achieves the
highest individual pass rate (\miVC{}), followed by flex test
execution (\miFlex{}) and ground-truth alignment
(\miGT{}). When requiring all three
graders to pass simultaneously (strict conjunction), \miAllThree{} of
tasks succeed.

\begin{table}[t]
\caption{Specification strategy comparison on the \datasetname{} dataset.
Regeneration score is on a 0--6 scale; per-grader rates are percentages.
Human-authored summary = diff title + description + test plan used
verbatim as the spec.}
\label{tab:human-baseline}
\centering
\begin{tabular}{@{}lcccc@{}}
\toprule
\textbf{Strategy} & \textbf{Regen} & \textbf{VC~\%} & \textbf{Flex~\%} & \textbf{GT~\%} \\
\midrule
Human-authored summary            & 4.23 & 64.0 & 60.0 & 50.0 \\
\toolname{}                       & 5.06 & 86.8 & 72.1 & 66.2 \\
\toolname{} + strengthening       & \textbf{\ssBest} & --   & --   & --   \\
\bottomrule
\end{tabular}
\end{table}

\subsection{RQ2: Spec Strengthening}
\label{sec:rq2}

{}

RQ2 asks whether per-task feedback from the three-tier verifier can
strengthen individual specifications that failed regeneration.
Starting from \ssTotal{} baseline specs, the strengthening
loop (Section~\ref{sec:refinement}) targets tasks with
regeneration score~$< 6.0$: \ssNull{} tasks are skipped due to null
baselines, leaving \ssConsidered{} considered tasks of which
\ssStrengthened{} enter strengthening (the remaining 37 already
score~$= 6.0$).
Each task undergoes up to \ssRounds{} rounds of refinement, with
per-task early exit once regeneration score reaches~6.0.

\begin{table}[t]
\caption{Spec strengthening round progression (Section~\ref{sec:rq2}, \ssTotal{}~tasks,
\ssRounds{}~rounds). Per-round metrics are over the \emph{failing
cohort} evaluated that round; the best-of row aggregates over all
\ssConsidered{} considered tasks.}
\label{tab:strengthening}
\centering
\begin{tabular}{@{}lrcccc@{}}
\toprule
\textbf{Round} & \textbf{Graded} & \textbf{Mean Regen} & \textbf{Flex~\%} & \textbf{VC~\%} & \textbf{GT~\%} \\
\midrule
Baseline & \ssConsidered & \ssBaseline & 72.1 & 86.8 & 66.2 \\
Round~1  & 31  & 4.68 & 60.0 & 86.7 & 54.8 \\
Round~2  & 19  & 4.71 & 55.6 & 88.9 & 52.6 \\
Round~3  & 11  & 4.38 & 45.5 & 81.8 & 45.5 \\
\midrule
\textbf{Best-of-K} & \ssConsidered & \textbf{\ssBest} & -- & -- & -- \\
\bottomrule
\end{tabular}
\end{table}

Table~\ref{tab:strengthening} shows the round-by-round progression.
Per-round regeneration scores are computed over only the failing tasks
evaluated that round (not all \ssConsidered{} tasks), so they are
naturally lower than the all-task baseline---the two are not directly
comparable.
The best-of-K row reports the aggregate after keeping the highest-scoring
spec per task across baseline and all rounds: the mean regeneration score
rises from \ssBaseline{} to \ssBest{} (\ssDelta{} delta) over the
\ssConsidered{} considered tasks.
Of the \ssStrengthened{} tasks that entered strengthening,
\ssImproved{} achieved a positive delta---an 84\% success rate.

\begin{figure}[t]
\centering
\includegraphics[width=\columnwidth]{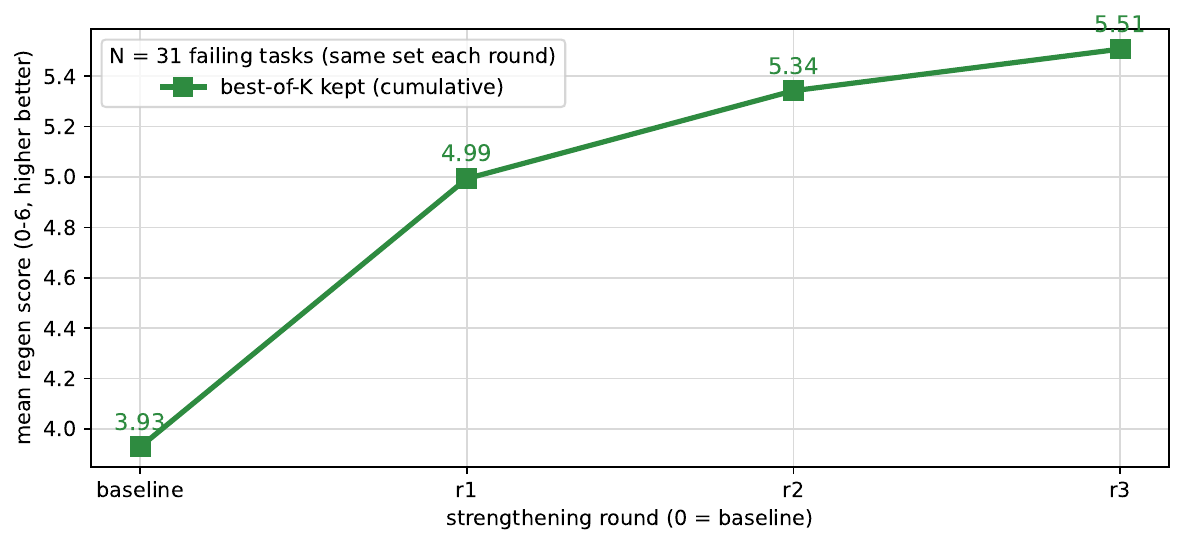}
\caption{Cumulative best-of-K regeneration score across \ssRounds{} strengthening rounds, over the failing cohort (same tasks at every point).}
\label{fig:strengthen-progression}
\end{figure}

Figure~\ref{fig:strengthen-progression} tracks the cumulative best-of-K
regeneration score over the failing cohort ($N=31$, the same tasks at every
round): it climbs steadily from a baseline of 3.93 to 4.99 after round~1,
5.34 after round~2, and 5.51 after round~3, showing that iterative
strengthening progressively recovers the failing cohort.

At the per-task level, \ssImproved{} of \ssStrengthened{} tasks
improve at least once across the three rounds, with many jumping
to $\geq 5.5$.

The graded-task count shrinks from 31 in round~1 to 19 in round~2
and 11 in round~3 as tasks that reach $\tau$ early-exit the pool.
VC pass rates remain high across rounds (82--89\%), while flex-test
and GT rates decline in later rounds (flex 60.0\% to 45.5\%, GT 54.8\%
to 45.5\%), reflecting the increasing difficulty of the residual
failing cohort.

Strengthening specs in this manner is an instance of \textit{loop engineering},
where a human gives an optimization target to an agent and let's it iterate
towards a goal.

\subsection{RQ3: Abstraction vs.\ Strength}
\label{sec:rq3}

{}

RQ3 investigates the tension between abstraction and strength: does a
more abstract spec---one that captures intent without implementation
detail---sacrifice regeneration fidelity?
For each task, we compute the compression ratio
$\text{CompressionRatio}(S, C) = |C|_\text{chars} / |S|_\text{chars}$,
where $|C|_\text{chars}$ is the character length of the ground-truth unified
diff and $|S|_\text{chars}$ is the character length of the generated spec.
Higher ratios indicate stronger compression.

\paragraph{Three extraction strategies.}
We compare three spec-extraction configurations spanning the
abstraction--detail trade-off: (a) \emph{Standard}, the default \toolname{}
extraction with no length constraint (the Section~\ref{sec:rq1} baseline);
(b) \emph{Constrained (1000c)}, targeting $\sim$1{,}000 characters for
maximally abstract, intent-focused specs; and (c) \emph{Diff-preserving},
prompted to retain concrete implementation detail (file paths, identifiers,
code snippets), yielding specs that approach the original diff size.
Table~\ref{tab:compression-variants} reports the results.

\begin{table}[t]
\caption{Spec abstraction vs.\ regeneration quality across three
extraction strategies. Higher compression ratio = more abstract spec;
regeneration score on 0--6 scale. Leakage = fraction of the diff's
distinctive code tokens that appear verbatim in the spec.}
\label{tab:compression-variants}
\centering
\begin{tabular}{@{}lrrrr@{}}
\toprule
\textbf{Strategy} & \textbf{Spec Avg} & \textbf{Ratio} & \textbf{Regen} & \textbf{Leak~\%} \\
\midrule
Constrained (1000c)            & 2{,}117  & 16.16$\times$ & 4.81 & 23.7 \\
Standard                       & 6{,}960  & 5.69$\times$  & 5.06 & 38.6 \\
\textbf{Diff-preserving}       & 11{,}836 & 2.79$\times$  & \textbf{5.55} & 62.6 \\
\bottomrule
\end{tabular}
\end{table}

At first glance, more concrete specs achieve higher regeneration scores:
the diff-preserving variant (2.79$\times$ compression) scores
5.55/6.0, outperforming the standard extraction (5.06,
5.69$\times$) and the constrained variant (4.81, 16.16$\times$).
Taken at face value, this would suggest that abstractness always
hurts---but the leakage column tells a different story.

\paragraph{Leakage confound.}
The diff-preserving spec copies 62.6\% of the diff's distinctive
code tokens verbatim (identifiers, literals), compared to 38.6\%
for standard and 23.7\% for constrained.
A per-tier breakdown shows that the regeneration score advantage is
disproportionately driven by the ground-truth alignment grader
(89\% vs.\ 63--66\%), which rewards matching the reference
patch---exactly what high-leakage enables.
The behavioral graders also improve (VC: 91\% vs.\ 81--87\%;
flex: 87\% vs.\ 66--72\%), but the gap is smaller.
This suggests that much of the diff-preserving advantage reflects
literal code copying rather than better intent communication.
The standard extraction offers the best balance of abstractness
and strength: meaningful compression (5.69$\times$) with moderate
leakage (38.6\%) and strong behavioral verification (VC~87\%,
flex~72\%).

\begin{figure}[t]
\centering
\includegraphics[width=0.75\columnwidth]{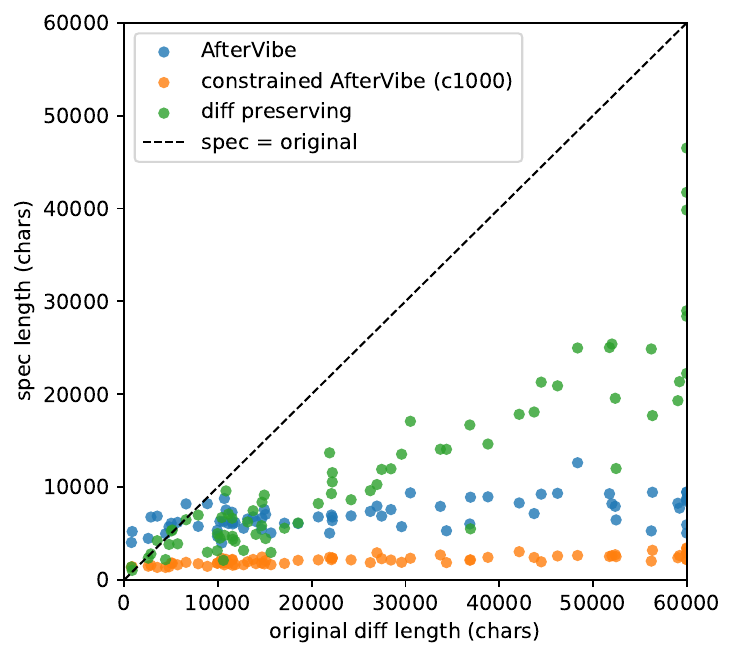}
\caption{Per-task spec length vs.\ original diff length for the three strategies (axes capped at 60K chars); the dashed line marks $\text{spec} = \text{original}$.}
\label{fig:orig-vs-spec}
\end{figure}

\paragraph{Abstractness confirmed by diversity.}
RQ4 provides complementary evidence: passing regenerations from the
standard spec are structurally diverse (chrF~0.89, not~1.0),
confirming that the spec constrains \emph{what} without dictating
\emph{how}---the hallmark of a well-abstracted specification.
The constrained strategy maintains tight spec lengths
($\sim$2{,}100~chars) regardless of diff size, while the
diff-preserving strategy scales with diff length
(Figure~\ref{fig:orig-vs-spec}).

\subsection{RQ4: Regeneration Diversity}
\label{sec:rq4}

{}

RQ4 measures how much independent regenerations vary when conditioned on the
same specification.
For each of the \miN{} tasks, we run the regeneration agent 3 times from the
same base commit and specification, yielding 200 task-pairs for pairwise
comparison (of the 216 possible pairs, 16 are skipped because one
rollout produced no patch).

\paragraph{Similarity metrics.}
We compare each pair of regenerated patches at the file level using two
complementary measures:
\emph{chrF}~(character $n$-gram F-score, normalized to $[0,1]$ via
\texttt{sacrebleu}) captures character-level overlap and is robust to minor
formatting differences;
\emph{difflib ratio}~(Python \texttt{SequenceMatcher}) captures
sequence-level structural similarity.
For each task-pair, we take the union of files touched by either patch;
files present on only one side score 0.0, and files present on both sides
are scored independently.
The task-level similarity is the mean of per-file scores.

\paragraph{Outcome segmentation.}
To relate code diversity to verification outcome, we segment task-pairs
by whether both regenerations pass (regeneration score~$\geq 5.5$), both fail,
or one passes and the other fails (``mixed''). This threshold matches
the pass-to-pass anchor below, so the both-pass segment and that anchor
cover the same 83 rollout pairs.
~
Across the 200 task-pairs, both-pass pairs show the highest code agreement
(chrF~0.888, difflib~0.819), confirming that passing specifications reliably
guide agents toward similar solutions.
Both-fail pairs remain nearly as similar (0.856/0.749)---under-specified specs
drive independent agents toward the \emph{same} incorrect solution, a signal of
systematic rather than random spec gaps---while mixed-outcome pairs are least
similar (0.835/0.740), consistent with genuine divergence driving the split
outcome (overall 0.858/0.771).

\paragraph{Cross-rollout stability.}
Across the 3 rollouts, the per-task regeneration score is stable (rollout
means of 5.06, 4.87, and 4.99; mean per-task standard deviation~0.49),
and within-task code agreement is high (both-pass chrF~0.89).
Independent regenerations from the same spec thus converge on
functionally equivalent solutions while still varying in structure.

\paragraph{Reference anchors (patch-level).}
To calibrate the similarity values, we compare rollout-pair similarity
against two reference points using added-lines-only patches:
(i)~\emph{upper bound}---each patch vs.\ itself with one added line
removed (mean chrF~0.982);
(ii)~\emph{lower bound}---cross-task pairs of unrelated patches
(mean chrF~0.180).
Pass-to-pass rollout pairs (both regeneration score~$\geq 5.5$) achieve
chrF~0.899, sitting much closer to the upper bound than the lower
bound.
This confirms that the specification constrains the solution space
far beyond what would be expected by chance, while still permitting
minor structural variation.

\subsection{RQ5: Grader Validity}
\label{sec:rq5}

{}
{}

RQ5 asks whether the extracted verification conditions and flex tests
genuinely discriminate correct code from incorrect code---that is, whether
they are valid oracles rather than superficial checks that pass on any
input.
We answer with three \emph{spec-free negative controls}: the same
verification conditions and test commands used in ground-truth validation
(Section~\ref{sec:dataset-cip}) are run against \emph{degraded},
\emph{pre-change}, or \emph{wrong} code instead of the original landed code.
No specification or regeneration agent is involved; only the \emph{code}
varies, so any difference in grader pass rate directly measures
discriminative power.

\paragraph{Pre-change control (before the change).}
Each task keeps its own oracle (verification conditions and test commands)
but is graded against the repository state
\emph{before} the task's code was landed (the parent commit).
If task~$A$'s VCs still pass on the pre-change checkout, the VCs are
not checking the task's intended behavior.

\paragraph{Degradation control (partial code).}
An LLM removes approximately 50\% of each task's diff hunks
(whole hunks, syntax-preserving), then the task's own VCs and tests
run against the degraded code.
If VCs do not notice the missing functionality, they lack coverage.

\paragraph{Permutation control (wrong code).}
Each task~$A$ keeps its own oracle but is graded against a
\emph{different} task~$B$'s landed code, assigned by a seeded
derangement (Sattolo's algorithm---a single-cycle permutation with
no fixed points, so every task receives unrelated code).
If task~$A$'s VCs still pass on task~$B$'s code, the VCs are not
checking task-specific intent.

\begin{table}[t]
\caption{Negative control results (Section~\ref{sec:rq5}), conditioned on the \datasetname{} tasks
whose correct code passes \emph{both} VC and flex (ground truth =
100\% by construction). Lower pass rates on worse code indicate
stronger discriminative power.}
\label{tab:negative-controls}
\centering
\begin{tabular}{@{}lrcc@{}}
\toprule
\textbf{Code quality} & \textbf{Jobs} & \textbf{VC~\%} & \textbf{Flex~\%} \\
\midrule
Correct (ground truth)        & 72  & 100.0\% & 100.0\% \\
Partial (50\% hunks removed)  & 66  & 47.0\%  & 42.4\%  \\
Pre-change (before change)    & 69  & 0.0\%   & 58.0\%  \\
Wrong (permutation)           & 70  & 1.4\%   & 51.4\%  \\
\bottomrule
\end{tabular}
\end{table}

\begin{figure}[t]
\centering
\includegraphics[width=\columnwidth]{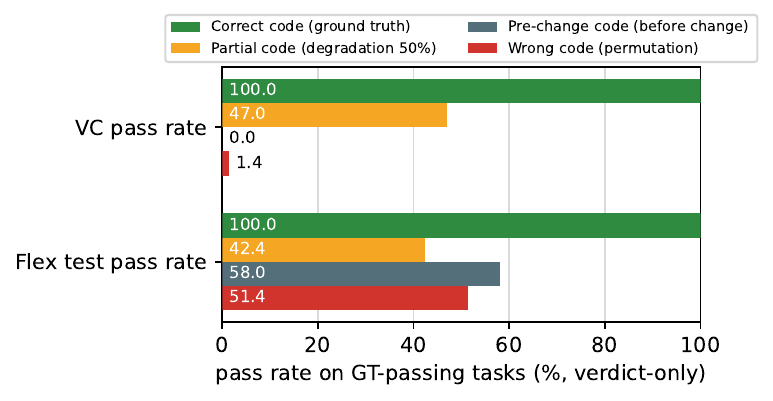}
\caption{Per-grader pass rates by code condition, conditioned on the 72
tasks whose correct code passes \emph{both} VC and flex.}
\label{fig:grader-by-condition}
\end{figure}

Table~\ref{tab:negative-controls} reports the conditioned view:
restricting to 72 tasks where correct code passes both graders
(ground truth = 100\% by construction), we measure how often each
grader \emph{also} passes degraded, pre-change, or wrong code.
The verification-clues grader exhibits sharp monotonic discrimination:
100\% on correct code $\to$ 47\% on partial $\to$ near-zero on
pre-change (0\%) and wrong code (1.4\%).
Not a single task's verification conditions pass on the repository
state before the change was landed, and only 1 of 70 passes on an
unrelated task's code, confirming that VCs check task-specific
behavioral intent rather than surface properties.
~
The flex-test grader drops to 42\% on partial code but remains
elevated on pre-change (58\%) and wrong code (51\%)---because many
test commands pre-exist the change and can pass on unrelated code
that happens to compile (e.g., ``build the target'').
~
Figure~\ref{fig:grader-by-condition} visualizes the conditioned
pass rates: VC's sharp staircase (100\% $\to$ 47\% $\to$
0--1\%) confirms strong task-specific discrimination.
~
These results validate the ground-truth filtering of
Section~\ref{sec:dataset-cip}: tasks whose VCs pass on correct code
are genuine positive signals, not false positives from overly
permissive oracles.

\subsection{RQ6: Comparison with the Human-Authored Summary}
\label{sec:rq6}

{}
{}

RQ6 asks whether the structured extraction of \toolname{} produces
specifications that are more effective for regeneration than the
descriptions developers already author as part of their normal
workflow.
We compare against a \emph{human-authored baseline}: the
human-authored summary that the developer submits for code review---consisting of
the diff title, description, and test plan---used verbatim as the
specification.
The regeneration agent, three-tier grading pipeline, and evaluation
infrastructure remain identical across both conditions, isolating
the effect of the specification source.

Table~\ref{tab:human-baseline} includes the comparison.
\toolname{} achieves a mean regeneration score of
5.06, compared to 4.23 for the human-authored summary---a difference
of 0.82 points on the 0--6 scale, with a consistent gap across all three
graders.\footnote{Robust to weighting: under equal $(2,2,2)$, swapped
$(2,3,1)$, and GT-heavy $(1,2,3)$ weights, the \toolname{}-vs-human gap stays
$0.82$--$0.89$ points and the per-task ranking is unchanged (Spearman~$\geq 0.97$).}
The human-authored summary is shorter on average
(2{,}129 vs.\ 6{,}960 characters), yet captures less of the intent
required for successful regeneration.

This result is consistent with the intended audience of each
artifact.
The human-authored summary is written for human reviewers who have access to
the accompanying code; it describes \emph{what changed} at a high
level but typically omits the behavioral constraints, design
decisions, and undiscoverable facts that a blind regeneration agent
requires.
\toolname's structured extraction recovers precisely these
elements from the conversation trajectory, producing specifications
that are more verbose but substantially more effective as
self-contained implementation guides.

\subsection{Implication of Results}
\label{sec:discussion}

Our results offer empirical support for the environmental grounding
hypothesis (Section~\ref{sec:grounding}).
The high pass rate after refinement suggests
that agents already extract substantial context from the repository
environment---build configurations, existing module structure, naming
conventions---without this information appearing in the spec. This has a
provocative implication: much of what we traditionally consider
``specification'' is actually shared knowledge that a grounded agent can
infer from the codebase. The \emph{useful} part of a specification---the
part that needs to be explicitly written---consists primarily of
project-specific decisions that cannot be derived from the environment.
This perspective clarifies the abstractness goal of
Section~\ref{sec:problem}: a spec should be concrete only for facts
the agent cannot discover from the environment, and abstract for
everything else.

As mentioned earlier, our vision is that such specifications can
supplant code as the artifact of human review and even the source of record;
exploring these possibilities in industrial practice is future work.

\section{Threats to Validity}
\label{sec:threats}

\textbf{Internal validity.}
Our oracle is LLM-generated, so the test suite could encode hallucinated or trivial behavior. We address this by validating every generated test against the original landed code, ensuring the suite encodes actual---not hallucinated---behavior; the negative-control analysis of Section~\ref{sec:rq5} further shows that the verification conditions sharply separate correct code from degraded, pre-change, and unrelated code. A second threat is reconstruction variance: because agents are non-deterministic, regenerating from the same specification can yield different \texttt{regen\_score}s across runs. We quantify this through the cross-rollout diversity analysis of Section~\ref{sec:rq4}, which shows that independent regenerations remain functionally stable (per-task \texttt{regen\_score} standard deviation of $0.49$) while varying in surface structure. Finally, there is a risk of information leakage between spec extraction and regeneration: if a single agent extracted the specification and then regenerated the code immediately afterward, it could still hold the original code in its context and reproduce it from memory rather than from the specification, inflating scores. We address this by separating the two stages into distinct sessions with no shared context, so the regenerating agent sees only the specification and cannot fall back on the original code.

\textbf{External validity.}
Our dataset consists of \miN{} tasks drawn from coding-assistant sessions, which may not be representative of all vibe-coded software or generalize to other languages, domains, or developer populations; we mitigate this by sampling tasks from real coding-assistant sessions rather than synthetic benchmarks, grounding the evaluation in genuine developer activity. In addition, all experiments use a single frontier LLM, and because specification quality depends on model capability, our results may not generalize to other or weaker LLMs.

\section{Related Work}
\label{sec:related}

\toolname\ draws on several lines of research concerned with specification extraction and its uses.

\subsection{Spec-Driven Development}

Recent industry tools such as GitHub SpecKit~\cite{github2025speckit} and Amazon Kiro~\cite{aws2025kiro}
have recently adopted spec-first workflows, requiring developers to author specifications upfront
and maintain them alongside code.
All of these approaches---from Design by Contract to modern spec-first IDEs---assume that
specifications are written before or during coding.
\toolname\ delivers similar benefits post-hoc, recovering specifications from sessions where
none were written, making it compatible with vibe coding by construction.
Kiro in particular uses the \emph{EARS} notation~\cite{mavinEasyApproachRequirements2009} (Easy Approach to Requirements Syntax) for its \texttt{requirements.md}, a small set of semi-formal templates that constrain otherwise unstructured natural-language requirements.
\toolname's verification conditions are themselves natural-language rubrics, and recent LLM-based methods have shown that such requirements can be autoformalized into EARS or related semi-formal templates~\cite{imranAutomatedEARSBasedRequirements2025, rosskothenConvertingNaturalLanguage2026, giannakopoulouAutomatedFormalizationStructured2021}.
\toolname\ could therefore act as an \emph{entry point} from vibe coding into spec-driven development.
Recasting code changes into natural language is also the goal of code summarization and
commit-message generation~\cite{liuNeuralmachinetranslationbasedCommitMessage2018, liuATOMCommitMessage2022, fengCodeBERTPreTrainedModel2020};
our specifications differ in that they must be precise enough to \emph{regenerate} the change, not merely describe it.

Traditional approaches, too, require writing specifications before or during implementation.
Meyer's Design by Contract~\cite{meyerApplyingDesignContract1992} mandates pre- and postconditions
written before code, with runtime checking enforcing compliance.
Test-driven development encodes intent as executable specifications that precede the code under test.
More recently, TiCoder~\cite{fakhouryLLMBasedTestDrivenInteractive2024} formalizes user intent interactively
during LLM-based code generation through iterative test refinement.

\subsection{LLM-Based Specification Extraction}

Close technical relatives of \toolname\ are systems that extract specifications from code.
SpecGen~\cite{maSpecGenAutomatedGeneration2025} generates formal JML specifications from Java methods
using LLMs combined with mutation testing and static analysis to filter incorrect candidates.
AutoSpec~\cite{wenEnchantingProgramSpecification2024} produces Dafny specifications through LLM-driven
counterexample-guided refinement. Endres et al.~\cite{endresCanLargeLanguage2024} generate postconditions
from natural-language docstrings, validated against existing test suites. SpecRover~\cite{ruanSpecRoverCodeIntent2025}
iteratively extracts code intent from buggy code and issue descriptions to guide automated patch generation.
All of these systems derive specifications from code or domain knowledge alone.
\toolname\ differs in two ways: it extracts from the agent trajectory---conversation and diff together---using
the conversation as a richer signal of developer intent; and it produces natural-language specs validated through
regeneration rather than formal annotations tied to a specific verification framework.

\subsection{Specification Verification and Quality}

A fundamental challenge in specifications is knowing whether the spec is correct.
VeriAct~\cite{misuVeriActVerifiabilityAgentic2026} shows that verifier-accepted Dafny specs are often
semantically wrong, and Le~Cong et al.~\cite{le-congCanLLMsReason2025} question whether LLM-generated
specs capture true semantics or merely surface patterns.
Lahiri frames intent formalization as a grand challenge~\cite{lahiriIntentFormalizationGrand2026} and
proposes grounding spec correctness in concrete input-output examples~\cite{lahiriEvaluatingLLMdrivenUserIntent2024}.
On the formal side, AutoVerus~\cite{yangAutoVerusAutomatedProof2025} generates machine-checked proofs
of Rust code from specifications, and Verus-SpecGym~\cite{agarwalVerusSpecGymAgenticEnvironment2026}
benchmarks natural-language-to-formal autoformalization---notably finding that execution-based checking
catches spec errors that LLM-as-judge evaluation misses.

These approaches rely on formal verification, symbolic checking, or executable formal specifications.
\toolname's answer is pragmatic: a spec is correct if it can regenerate functionally equivalent code,
validated by a multi-tier pipeline---no theorem proving required.
It works in the opposite direction, recovering natural-language specs post-hoc rather than formalizing
intent up front, and thus operationalizes Lahiri's vision without formal specification languages.
The Kiro team~\cite{Kiro2026DeepSpecAnalysis} similarly use LLM reasoning to elicit tests from informal
requirements, echoing our extraction of verification conditions from conversations.

\subsection{Agent Trajectories as Artifacts}

An emerging body of work uses or produces trajectory data from coding agents.
Zhu et al.~\cite{zhuSpecificationVibingAutomated2026} generate specifications from failing tests
and natural-language descriptions, then use them to guide program repair. RECODE-H~\cite{miaoRECODEHBenchmarkResearch2025}
benchmarks iterative code generation with multi-level human feedback, treating the full interaction as a trajectory.
TrajAudit~\cite{wangTrajAuditAutomatedFailure2026} also operates on agent trajectories, but for failure
diagnosis---localizing where and why an agent went wrong rather than capturing what it achieved.

These works use trajectories to generate code, benchmark interactions, or diagnose failures.
\toolname\ instead treats trajectories as a specification source, extracting self-validating
verification conditions and regeneration specifications from the conversation record.
\toolname\ is, to our knowledge, the first system to use agent coding trajectories---conversation and diff
together---for post-hoc intent capture.

\section{Conclusion}
\label{sec:conclusion}

We have presented \toolname, a framework for distilling verifiable
specifications from vibe-coded artifacts. By extracting structured
specs and validating them through a regeneration test backed by a three-tier
verification pipeline---flexible test execution, independent verification
conditions, and ground-truth alignment---\toolname\ bridges the gap
between the ease of vibe coding and the rigor of traditional software
engineering. Our
evaluation measures whether extracted specs can regenerate ground-truth diffs,
how the level of abstraction relates to regeneration fidelity, how much
independent regenerations vary, and whether feedback can strengthen
individual specs.
We believe that specifications---not code---should become the primary
artifact of human authorship in the age of AI-assisted development, and
\toolname\ is a first step toward realizing this vision.

\section*{Data Availability}
\label{sec:data-availability}

We provide the full pipeline prompts as supplementary material for
reproducibility---and, to our knowledge, no human
vibe-coding trajectory dataset with execution environments is yet openly
available.

\section*{Acknowledgments}

Use of generative AI.
We disclose that generative AI was used throughout this work---drafting
text, writing analysis scripts, processing data, and generating
tables---with all output reviewed and verified by the authors, who take
full responsibility for the final manuscript.

\bibliographystyle{plainnat}
\bibliography{MatteoAtMeta}

@misc{agarwalVerusSpecGymAgenticEnvironment2026,
  title = {Verus-{{SpecGym}}: {{An Agentic Environment}} for {{Evaluating Specification Autoformalization}}},
  shorttitle = {Verus-{{SpecGym}}},
  author = {Agarwal, Anmol and Neamtu, Natalie and Aggarwal, Pranjal and Kim, Seungone and Limperg, Jannis and Flamant, Cedric and Shimizu, Kanna and Parno, Bryan and Welleck, Sean},
  year = 2026,
  month = may,
  journal = {arXiv.org},
  urldate = {2026-06-30},
  howpublished = {https://arxiv.org/abs/2605.26457v1},
  langid = {english}
}

@misc{aws2025kiro,
  title = {Kiro: {{Move}} beyond {{AI}} Coding to Agentic Engineering},
  author = {{Amazon Web Services}},
  year = 2025
}

@inproceedings{bacchelliExpectationsOutcomesChallenges2013,
  title = {Expectations, Outcomes, and Challenges of Modern Code Review},
  booktitle = {2013 35th {{International Conference}} on {{Software Engineering}} ({{ICSE}})},
  author = {Bacchelli, Alberto and Bird, Christian},
  year = 2013,
  month = may,
  pages = {712--721},
  issn = {1558-1225},
  doi = {10.1109/ICSE.2013.6606617},
  urldate = {2026-06-25}
}

@inproceedings{chen1978nversion,
  title = {N-Version Programming: A Fault-Tolerance Approach to Reliability of Software Operation},
  booktitle = {Digest of Papers, {{FTCS-8}}: {{Eighth}} Annual International Conference on Fault-Tolerant Computing},
  author = {Chen, Liming and Avizienis, Algirdas},
  year = 1978,
  pages = {3--9},
  address = {Toulouse, France}
}

@article{endresCanLargeLanguage2024,
  title = {Can {{Large Language Models Transform Natural Language Intent}} into {{Formal Method Postconditions}}?},
  author = {Endres, Madeline and Fakhoury, Sarah and Chakraborty, Saikat and Lahiri, Shuvendu K.},
  year = 2024,
  month = jul,
  journal = {Proceedings of the ACM on Software Engineering},
  volume = {1},
  number = {FSE},
  pages = {84:1889--84:1912},
  doi = {10.1145/3660791},
  urldate = {2026-06-15}
}

@article{fakhouryLLMBasedTestDrivenInteractive2024,
  title = {{{LLM-Based Test-Driven Interactive Code Generation}}: {{User Study}} and {{Empirical Evaluation}}},
  shorttitle = {{{LLM-Based Test-Driven Interactive Code Generation}}},
  author = {Fakhoury, Sarah and Naik, Aaditya and Sakkas, Georgios and Chakraborty, Saikat and Lahiri, Shuvendu K.},
  year = 2024,
  month = sep,
  journal = {IEEE Transactions on Software Engineering},
  volume = {50},
  number = {9},
  pages = {2254--2268},
  issn = {1939-3520},
  doi = {10.1109/TSE.2024.3428972},
  urldate = {2026-06-15}
}

@inproceedings{fengCodeBERTPreTrainedModel2020,
  title = {{{CodeBERT}}: {{A Pre-Trained Model}} for {{Programming}} and {{Natural Languages}}},
  shorttitle = {{{CodeBERT}}},
  booktitle = {Findings of the {{Association}} for {{Computational Linguistics}}: {{EMNLP}} 2020},
  author = {Feng, Zhangyin and Guo, Daya and Tang, Duyu and Duan, Nan and Feng, Xiaocheng and Gong, Ming and Shou, Linjun and Qin, Bing and Liu, Ting and Jiang, Daxin and Zhou, Ming},
  editor = {Cohn, Trevor and He, Yulan and Liu, Yang},
  year = 2020,
  month = nov,
  pages = {1536--1547},
  publisher = {Association for Computational Linguistics},
  address = {Online},
  doi = {10.18653/v1/2020.findings-emnlp.139},
  urldate = {2026-06-30}
}

@misc{Fuzz4AllUniversalFuzzing,
  title = {{{Fuzz4All}}: {{Universal Fuzzing}} with {{Large Language Models}} \textbar{} {{Proceedings}} of the {{IEEE}}/{{ACM}} 46th {{International Conference}} on {{Software Engineering}}},
  urldate = {2026-06-30},
  howpublished = {https://dl.acm.org/doi/10.1145/3597503.3639121}
}

@misc{github2025speckit,
  title = {Spec Kit: {{Toolkit}} for Spec-Driven Development},
  author = {{GitHub}},
  year = 2025
}

@inproceedings{gotelAnalysisRequirementsTraceability1994,
  title = {An Analysis of the Requirements Traceability Problem},
  booktitle = {Proceedings of {{IEEE International Conference}} on {{Requirements Engineering}}},
  author = {Gotel, O.C.Z. and Finkelstein, C.W.},
  year = 1994,
  month = apr,
  pages = {94--101},
  doi = {10.1109/ICRE.1994.292398},
  urldate = {2026-06-30}
}

@misc{guSurveyLLMasaJudge2025,
  title = {A {{Survey}} on {{LLM-as-a-Judge}}},
  author = {Gu, Jiawei and Jiang, Xuhui and Shi, Zhichao and Tan, Hexiang and Zhai, Xuehao and Xu, Chengjin and Li, Wei and Shen, Yinghan and Ma, Shengjie and Liu, Honghao and Wang, Saizhuo and Zhang, Kun and Wang, Yuanzhuo and Gao, Wen and Ni, Lionel and Guo, Jian},
  year = 2025,
  month = oct,
  number = {arXiv:2411.15594},
  eprint = {2411.15594},
  primaryclass = {cs},
  publisher = {arXiv},
  doi = {10.48550/arXiv.2411.15594},
  urldate = {2026-02-04},
  archiveprefix = {arXiv}
}

@article{knightExperimentalEvaluationAssumption1986,
  title = {An Experimental Evaluation of the Assumption of Independence in Multiversion Programming},
  author = {Knight, John C. and Leveson, Nancy G.},
  year = 1986,
  month = jan,
  journal = {IEEE Transactions on Software Engineering},
  volume = {SE-12},
  number = {1},
  pages = {96--109},
  issn = {1939-3520},
  doi = {10.1109/TSE.1986.6312924},
  urldate = {2026-06-30}
}

@inproceedings{lahiriEvaluatingLLMdrivenUserIntent2024,
  title = {Evaluating {{LLM-driven User-Intent Formalization}} for {{Verification-Aware Languages}}},
  booktitle = {Proceedings of the 24th {{Conference}} on {{Formal Methods}} in {{Computer-Aided Design}} -- {{FMCAD}} 2024},
  author = {Lahiri, Shuvendu K.},
  year = 2024,
  month = oct,
  pages = {142--147},
  publisher = {TU Wien Academic Press},
  doi = {10.34727/2024/isbn.978-3-85448-065-5_19},
  urldate = {2026-06-26},
  copyright = {https://creativecommons.org/licenses/by/4.0/},
  isbn = {978-3-85448-065-5},
  langid = {english}
}

@misc{lahiriIntentFormalizationGrand2026,
  title = {Intent {{Formalization}}: {{A Grand Challenge}} for {{Reliable Coding}} in the {{Age}} of {{AI Agents}}},
  shorttitle = {Intent {{Formalization}}},
  author = {Lahiri, Shuvendu K.},
  year = 2026,
  month = mar,
  number = {arXiv:2603.17150},
  eprint = {2603.17150},
  primaryclass = {cs.SE},
  publisher = {arXiv},
  doi = {10.48550/arXiv.2603.17150},
  urldate = {2026-06-15},
  archiveprefix = {arXiv}
}

@misc{le-congCanLLMsReason2025,
  title = {Can {{LLMs Reason About Program Semantics}}? {{A Comprehensive Evaluation}} of {{LLMs}} on {{Formal Specification Inference}}},
  shorttitle = {Can {{LLMs Reason About Program Semantics}}?},
  author = {{Le-Cong}, Thanh and Le, Bach and Murray, Toby},
  year = 2025,
  month = may,
  number = {arXiv:2503.04779},
  eprint = {2503.04779},
  primaryclass = {cs.PL},
  publisher = {arXiv},
  doi = {10.48550/arXiv.2503.04779},
  urldate = {2026-06-15},
  archiveprefix = {arXiv}
}

@article{liuATOMCommitMessage2022,
  title = {{{ATOM}}: {{Commit Message Generation Based}} on {{Abstract Syntax Tree}} and {{Hybrid Ranking}}},
  shorttitle = {{{ATOM}}},
  author = {Liu, Shangqing and Gao, Cuiyun and Chen, Sen and Nie, Lun Yiu and Liu, Yang},
  year = 2022,
  month = may,
  journal = {IEEE Transactions on Software Engineering},
  volume = {48},
  number = {5},
  pages = {1800--1817},
  issn = {1939-3520},
  doi = {10.1109/TSE.2020.3038681},
  urldate = {2026-06-30}
}

@inproceedings{liuNeuralmachinetranslationbasedCommitMessage2018,
  title = {Neural-Machine-Translation-Based Commit Message Generation: How Far Are We?},
  shorttitle = {Neural-Machine-Translation-Based Commit Message Generation},
  booktitle = {Proceedings of the 33rd {{ACM}}/{{IEEE International Conference}} on {{Automated Software Engineering}}},
  author = {Liu, Zhongxin and Xia, Xin and Hassan, Ahmed E. and Lo, David and Xing, Zhenchang and Wang, Xinyu},
  year = 2018,
  month = sep,
  series = {{{ASE}} '18},
  pages = {373--384},
  publisher = {Association for Computing Machinery},
  address = {New York, NY, USA},
  doi = {10.1145/3238147.3238190},
  urldate = {2026-06-30},
  isbn = {978-1-4503-5937-5}
}

@inproceedings{madaanSELFREFINEIterativeRefinement2023,
  title = {{{SELF-REFINE}}: Iterative Refinement with Self-Feedback},
  shorttitle = {{{SELF-REFINE}}},
  booktitle = {Proceedings of the 37th {{International Conference}} on {{Neural Information Processing Systems}}},
  author = {Madaan, Aman and Tandon, Niket and Gupta, Prakhar and Hallinan, Skyler and Gao, Luyu and Wiegreffe, Sarah and Alon, Uri and Dziri, Nouha and Prabhumoye, Shrimai and Yang, Yiming and Gupta, Shashank and Majumder, Bodhisattwa Prasad and Hermann, Katherine and Welleck, Sean and Yazdanbakhsh, Amir and Clark, Peter},
  year = 2023,
  month = dec,
  series = {{{NIPS}} '23},
  pages = {46534--46594},
  publisher = {Curran Associates Inc.},
  address = {Red Hook, NY, USA},
  urldate = {2026-06-29}
}

@inproceedings{maSpecGenAutomatedGeneration2025,
  title = {{{SpecGen}}: {{Automated Generation}} of {{Formal Program Specifications}} via {{Large Language Models}}},
  shorttitle = {{{SpecGen}}},
  booktitle = {2025 {{IEEE}}/{{ACM}} 47th {{International Conference}} on {{Software Engineering}} ({{ICSE}})},
  author = {Ma, Lezhi and Liu, Shangqing and Li, Yi and Xie, Xiaofei and Bu, Lei},
  year = 2025,
  month = apr,
  pages = {16--28},
  issn = {1558-1225},
  doi = {10.1109/ICSE55347.2025.00129},
  urldate = {2026-06-15}
}

@article{meyerApplyingDesignContract1992,
  title = {Applying 'Design by Contract'},
  author = {Meyer, B.},
  year = 1992,
  month = oct,
  journal = {Computer},
  volume = {25},
  number = {10},
  pages = {40--51},
  issn = {1558-0814},
  doi = {10.1109/2.161279},
  urldate = {2026-06-15}
}

@misc{miaoRECODEHBenchmarkResearch2025,
  title = {{{RECODE-H}}: {{A Benchmark}} for {{Research Code Development}} with {{Interactive Human Feedback}}},
  shorttitle = {{{RECODE-H}}},
  author = {Miao, Chunyu and Zou, Henry Peng and Li, Yangning and Chen, Yankai and Wang, Yibo and Wang, Fangxin and Li, Yifan and Yang, Wooseong and He, Bowei and Zhang, Xinni and Yu, Dianzhi and Yang, Hanchen and Nguyen, Hoang H. and Zhou, Yue and Yang, Jie and Guo, Jizhou and Fan, Wenzhe and Yeh, Chin-Yuan and Meng, Panpan and Fang, Liancheng and Qi, Jinhu and Huang, Wei-Chieh and Gu, Zhengyao and Han, Yuwei and He, Langzhou and Yang, Yuyao and Li, Yinghui and Zheng, Hai-Tao and Liu, Xue and King, Irwin and Yu, Philip S.},
  year = 2025,
  month = oct,
  number = {arXiv:2510.06186},
  eprint = {2510.06186},
  primaryclass = {cs.CL},
  publisher = {arXiv},
  doi = {10.48550/arXiv.2510.06186},
  urldate = {2026-06-15},
  archiveprefix = {arXiv}
}

@misc{misuVeriActVerifiabilityAgentic2026,
  title = {{{VeriAct}}: {{Beyond Verifiability}} -- {{Agentic Synthesis}} of {{Correct}} and {{Complete Formal Specifications}}},
  shorttitle = {{{VeriAct}}},
  author = {Misu, Md Rakib Hossain and Ma, Iris and Lopes, Cristina V.},
  year = 2026,
  month = mar,
  number = {arXiv:2604.00280},
  eprint = {2604.00280},
  primaryclass = {cs.SE},
  publisher = {arXiv},
  doi = {10.48550/arXiv.2604.00280},
  urldate = {2026-06-15},
  archiveprefix = {arXiv}
}

@inproceedings{ruanSpecRoverCodeIntent2025,
  title = {{{SpecRover}}: {{Code Intent Extraction}} via {{LLMs}}},
  shorttitle = {{{SpecRover}}},
  booktitle = {2025 {{IEEE}}/{{ACM}} 47th {{International Conference}} on {{Software Engineering}} ({{ICSE}})},
  author = {Ruan, Haifeng and Zhang, Yuntong and Roychoudhury, Abhik},
  year = 2025,
  month = apr,
  pages = {963--974},
  issn = {1558-1225},
  doi = {10.1109/ICSE55347.2025.00080},
  urldate = {2026-06-15}
}

@misc{ugareAgenticCodeReasoning2026,
  title = {Agentic {{Code Reasoning}}},
  author = {Ugare, Shubham and Chandra, Satish},
  year = 2026,
  month = mar,
  number = {arXiv:2603.01896},
  eprint = {2603.01896},
  primaryclass = {cs.SE},
  publisher = {arXiv},
  doi = {10.48550/arXiv.2603.01896},
  urldate = {2026-06-01},
  archiveprefix = {arXiv}
}

@misc{wangTrajAuditAutomatedFailure2026,
  title = {{{TrajAudit}}: {{Automated Failure Diagnosis}} for {{Agentic Coding Systems}}},
  shorttitle = {{{TrajAudit}}},
  author = {Wang, Minxing and Xie, Xiaofei and Huo, Yintong},
  year = 2026,
  month = may,
  number = {arXiv:2605.26563},
  eprint = {2605.26563},
  primaryclass = {cs.SE},
  publisher = {arXiv},
  doi = {10.48550/arXiv.2605.26563},
  urldate = {2026-06-26},
  archiveprefix = {arXiv}
}

@inproceedings{wenEnchantingProgramSpecification2024,
  title = {Enchanting {{Program Specification Synthesis}} by~{{Large Language Models Using Static Analysis}} and~{{Program Verification}}},
  booktitle = {Computer {{Aided Verification}}},
  author = {Wen, Cheng and Cao, Jialun and Su, Jie and Xu, Zhiwu and Qin, Shengchao and He, Mengda and Li, Haokun and Cheung, Shing-Chi and Tian, Cong},
  editor = {Gurfinkel, Arie and Ganesh, Vijay},
  year = 2024,
  pages = {302--328},
  publisher = {Springer Nature Switzerland},
  address = {Cham},
  doi = {10.1007/978-3-031-65630-9_16},
  isbn = {978-3-031-65630-9},
  langid = {english}
}

@inproceedings{xiaAutomatedProgramRepair2024,
  title = {Automated {{Program Repair}} via {{Conversation}}: {{Fixing}} 162 out of 337 {{Bugs}} for \$0.42 {{Each}} Using {{ChatGPT}}},
  shorttitle = {Automated {{Program Repair}} via {{Conversation}}},
  booktitle = {Proceedings of the 33rd {{ACM SIGSOFT International Symposium}} on {{Software Testing}} and {{Analysis}}},
  author = {Xia, Chunqiu Steven and Zhang, Lingming},
  year = 2024,
  month = sep,
  series = {{{ISSTA}} 2024},
  pages = {819--831},
  publisher = {Association for Computing Machinery},
  address = {New York, NY, USA},
  doi = {10.1145/3650212.3680323},
  urldate = {2026-02-24},
  isbn = {979-8-4007-0612-7}
}

@article{yangAutoVerusAutomatedProof2025,
  title = {{{AutoVerus}}: {{Automated Proof Generation}} for {{Rust Code}}},
  shorttitle = {{{AutoVerus}}},
  author = {Yang, Chenyuan and Li, Xuheng and Misu, Md Rakib Hossain and Yao, Jianan and Cui, Weidong and Gong, Yeyun and Hawblitzel, Chris and Lahiri, Shuvendu and Lorch, Jacob R. and Lu, Shuai and Yang, Fan and Zhou, Ziqiao and Lu, Shan},
  year = 2025,
  month = oct,
  journal = {Proceedings of the ACM on Programming Languages},
  volume = {9},
  number = {OOPSLA2},
  pages = {396:3454--396:3482},
  doi = {10.1145/3763174},
  urldate = {2026-06-26}
}

@inproceedings{zhengJudgingLLMasajudgeMTbench2023,
  title = {Judging {{LLM-as-a-judge}} with {{MT-bench}} and {{Chatbot Arena}}},
  booktitle = {Proceedings of the 37th {{International Conference}} on {{Neural Information Processing Systems}}},
  author = {Zheng, Lianmin and Chiang, Wei-Lin and Sheng, Ying and Zhuang, Siyuan and Wu, Zhanghao and Zhuang, Yonghao and Lin, Zi and Li, Zhuohan and Li, Dacheng and Xing, Eric P. and Zhang, Hao and Gonzalez, Joseph E. and Stoica, Ion},
  year = 2023,
  month = dec,
  series = {{{NIPS}} '23},
  pages = {46595--46623},
  publisher = {Curran Associates Inc.},
  address = {Red Hook, NY, USA},
  urldate = {2026-03-16}
}

@inproceedings{zhugeAgentasaJudgeEvaluateAgents2025,
  title = {Agent-as-a-{{Judge}}: {{Evaluate Agents}} with {{Agents}}},
  shorttitle = {Agent-as-a-{{Judge}}},
  booktitle = {Proceedings of the 42nd {{International Conference}} on {{Machine Learning}}},
  author = {Zhuge, Mingchen and Zhao, Changsheng and Ashley, Dylan R. and Wang, Wenyi and Khizbullin, Dmitrii and Xiong, Yunyang and Liu, Zechun and Chang, Ernie and Krishnamoorthi, Raghuraman and Tian, Yuandong and Shi, Yangyang and Chandra, Vikas and Schmidhuber, J{\"u}rgen},
  year = 2025,
  month = oct,
  pages = {80569--80611},
  publisher = {PMLR},
  issn = {2640-3498},
  urldate = {2026-06-30},
  langid = {english}
}

@misc{zhuSpecificationVibingAutomated2026,
  title = {Specification {{Vibing}} for {{Automated Program Repair}}},
  author = {Zhu, Taohong and Cordeiro, Lucas C. and Mustafa, Mustafa A. and Sun, Youcheng},
  year = 2026,
  month = feb,
  number = {arXiv:2602.08263},
  eprint = {2602.08263},
  primaryclass = {cs.SE},
  publisher = {arXiv},
  doi = {10.48550/arXiv.2602.08263},
  urldate = {2026-06-15},
  archiveprefix = {arXiv}
}

@misc{Dora2025,
  title = {State of {{AI}}-Assisted Software: {{DORA}} Report},
  author = {{DORA}},
  year = 2025,
  howpublished = {\url{https://dora.dev/research/2025/dora-report/}},
  url = {https://dora.dev/research/2025/dora-report/},
  urldate = {2025-11-01},
  note = {Online; DORA (DevOps Research and Assessment), Google Cloud}
}

@misc{Kiro2026DeepSpecAnalysis,
  title = {Deep Dive: {{Spec}} Analysis and Requirements Verification in {{Kiro}}},
  author = {{Kiro Team}},
  year = 2026,
  howpublished = {Kiro Blog, \url{https://kiro.dev/blog/}},
  url = {https://kiro.dev/blog/},
  urldate = {2026-06-01},
  note = {Kiro Blog}
}

@inproceedings{mavinEasyApproachRequirements2009,
  title = {Easy {{Approach}} to {{Requirements Syntax}} ({{EARS}})},
  booktitle = {Proceedings of the 2009 17th {{IEEE International Requirements Engineering Conference}}, {{RE}}},
  author = {Mavin, Alistair and Wilkinson, Philip and Harwood, Adrian and Novak, Mark},
  year = 2009,
  month = aug,
  series = {{{RE}} '09},
  pages = {317--322},
  publisher = {IEEE Computer Society},
  address = {USA},
  doi = {10.1109/RE.2009.9},
  urldate = {2026-07-09},
  isbn = {978-0-7695-3761-0}
}

@inproceedings{imranAutomatedEARSBasedRequirements2025,
  title = {Automated {{EARS-Based Requirements Generation}} with {{Lightweight Large Language Models}}},
  booktitle = {2025 {{IEEE International Conference}} on {{Technology Management}}, {{Operations}} and {{Decisions}} ({{ICTMOD}})},
  author = {Imran, Muhammad Huzaifa and Tahir, Touseef and Hassan, Bilal and Jahankhani, Hamid},
  year = 2025,
  month = oct,
  pages = {1--6},
  issn = {2159-5119},
  doi = {10.1109/ICTMOD66732.2025.11371998},
  urldate = {2026-07-09}
}

@inproceedings{rosskothenConvertingNaturalLanguage2026,
  title = {{On Converting Natural Language Requirements into Semi-Formal Templates Using LLMs}},
  booktitle = {{2026 IEEE 34th International Requirements Engineering Conference (RE)}},
  author = {Ro{\ss}kothen, Julian and Fuch{\ss}, Dominik and Erd{\"o}si, Florian and Floru{\ss}, Maria and Keim, Jan and Hey, Tobias},
  year = 2026,
  urldate = {2026-07-09},
  langid = {ngerman}
}

@article{giannakopoulouAutomatedFormalizationStructured2021,
  title = {Automated Formalization of Structured Natural Language Requirements},
  author = {Giannakopoulou, Dimitra and Pressburger, Thomas and Mavridou, Anastasia and Schumann, Johann},
  year = 2021,
  month = sep,
  journal = {Information and Software Technology},
  volume = {137},
  pages = {106590},
  issn = {0950-5849},
  doi = {10.1016/j.infsof.2021.106590},
  urldate = {2026-07-09}
}

\definecolor{promptframe}{HTML}{2C3E50}   %
\definecolor{promptbg}{HTML}{F8FAFC}      %
\definecolor{promptrule}{HTML}{B7C2CC}    %
\definecolor{promptbreak}{HTML}{95A5A6}   %

\lstdefinestyle{promptstyle}{%
  basicstyle=\fontfamily{lmtt}\selectfont\footnotesize,
  breaklines=true,
  breakatwhitespace=true,
  breakindent=0pt,
  postbreak=\mbox{\textcolor{promptbreak}{$\hookrightarrow$}\space},
  columns=fixed,
  keepspaces=true,
  showstringspaces=false,
  upquote=true,
  extendedchars=true,
  tabsize=2,
  aboveskip=0pt,
  belowskip=0pt,
  xleftmargin=0pt,
  xrightmargin=0pt,
  literate={—}{{\textemdash}}1 {–}{{\textendash}}1,
}

\newtcolorbox[auto counter, number within=section,
  crefname={prompt}{prompts}, Crefname={Prompt}{Prompts}]{promptbox}[2][]{%
  breakable, enhanced,
  colback=promptbg,
  colframe=promptframe,
  colbacktitle=promptframe,
  coltitle=white,
  fonttitle=\bfseries\sffamily\small,
  fontupper=\normalfont,
  boxrule=0.6pt,
  arc=2.5pt,
  left=7pt, right=7pt, top=5pt, bottom=5pt,
  toptitle=2.5pt, bottomtitle=2.5pt,
  title={Prompt~\thetcbcounter\enspace\textendash\enspace #2},
  label={#1},
}

\newcommand{\promptartifact}[4]{%
  \begin{promptbox}[label={#1}]{#2}%
    {\itshape\small #3\par}%
    \vspace{2pt}%
    \noindent\textcolor{promptrule}{\rule{\linewidth}{0.4pt}}\par
    \vspace{3pt}%
    \lstinputlisting[style=promptstyle]{#4}%
  \end{promptbox}%
  \medskip
}

\clearpage
\appendix
\section{Prompt Artifacts}
\label{app:prompts}

This appendix reproduces, verbatim, the full prompt suite behind
\toolname{}'s post-hoc specification and verification pipeline
(\cref{sec:approach}), released as supplementary material
(\cref{sec:data-availability}). Each box is a single artifact read
directly from our released prompt package; template placeholders of the
form \texttt{\{...\}} are left unfilled so the exact prompt as run is
visible. The artifacts fall into two groups: the specification and
extraction prompts that distill a reusable specification, verification
conditions, and a runnable test manifest from a coding session
(\cref{app:prompts-pipeline}), and the multi-tier verification graders
that score code regenerated from a specification
(\cref{app:prompts-graders}).

\subsection{Specification and Extraction Pipeline}
\label{app:prompts-pipeline}

These five prompts run over a coding session (its conversation plus the
resulting diff) to produce the abstract specification and the evidence
used to verify code later regenerated from it.

\begin{promptbox}[label={app:prompt-spec-creation}]{Specification Creation}%
    {\itshape\small Distills a structured, abstract specification from a code change and its
   originating conversation so a blind coding agent can regenerate
   functionally equivalent code.\par}%
    \vspace{2pt}%
    \noindent\textcolor{promptrule}{\rule{\linewidth}{0.4pt}}\par
    \vspace{3pt}%
\begin{lstlisting}[style=promptstyle]
Given the following conversation and diff, create a specification that another coding agent can use to reimplement the change.

## Part 1: Specification
Write three short prose sections (no code, identifiers, paths, or line numbers unless essential and undiscoverable):
- **Intent and Rationale**: what the code changes in the conversation and diff achieve
- **Essential Design Decisions**: the key behaviors/criteria that must be preserved (separate essential decisions from incidental implementation choices).
- **Undiscoverable Facts**: external names, thresholds, contracts, and domain gotchas an implementer cannot infer from the codebase. This includes any exact file paths, filenames, or symbol names the user explicitly requested in the conversation — preserve those verbatim, since they are a required part of the contract.

## Part 2: Requirements
List the behavioral conditions a reviewer would check, each phrased as a declarative statement. Focus on behavior, not code structure. Avoid over-specificity that would match only one implementation.

Focus on OUTCOMES, not IMPLEMENTATION. Avoid:
- File names, function names, or variable names — UNLESS the user explicitly specified that exact name or path in the conversation, in which case it is a hard requirement: keep it verbatim and add a requirement stating that the named artifact exists and behaves as asked
- Implementation details or code snippets
- Line numbers or specific code locations
- Mentioning changes explicitly

The spec should allow someone to verify correctness WITHOUT reading the actual code.

## CRITICAL: Test File Handling
Do NOT include instructions to add, modify, or write tests, and do NOT add verification conditions about test coverage or tests passing. Test changes are already present in the working directory; focus only on the non-test production code.

Conversation context:
```
{conversation_context}
```

Unified Diff:
```
{unified_diff}
```
\end{lstlisting}%
  \end{promptbox}%
  \medskip

\begin{promptbox}[label={app:prompt-spec-strengthen}]{Per-Task Specification Strengthening}%
    {\itshape\small Rewrites a specification whose regenerated code scored below threshold,
   used in a best-of-$K$-over-rounds loop that keeps the highest-scoring
   specification per task.\par}%
    \vspace{2pt}%
    \noindent\textcolor{promptrule}{\rule{\linewidth}{0.4pt}}\par
    \vspace{3pt}%
\begin{lstlisting}[style=promptstyle]
You are refining a specification that failed the regeneration test. A blind coding agent was given only the specification below and asked to reimplement a code change from scratch; the regenerated code did not satisfy one of the behavioral verification conditions expected of the original change.

Your task is to analyze why the regeneration diverged from the original intent, identify what was ambiguous or underspecified in the specification, and produce a complete rewritten specification that closes that gap. Output the full rewritten specification — not a diff, not a list of edits — so it can be used on its own by a fresh agent that has never seen the original code or conversation.

## Inputs

ORIGINAL SPECIFICATION (S):
```
{original_spec}
```

GROUND-TRUTH CODE CHANGE (C):
```
{ground_truth_diff}
```

REGENERATED CODE CHANGE (C') produced from the specification above:
```
{regenerated_diff}
```

FAILED VERIFICATION CONDITION (a behavioral property the regeneration did not satisfy):
```
{failed_verification_condition}
```

GRADER'S REJECTION REASONING:
```
{grader_rejection_reasoning}
```

## Analysis instructions

1. Compare the ground-truth change (C) with the regenerated change (C') and pinpoint where the regeneration diverged from the intended behavior.
2. Identify what in the original specification (S) was ambiguous, missing, or underspecified such that a competent implementer produced the divergent behavior — focus on the failed verification condition and the grader's reasoning.
3. Rewrite the specification so that the failed behavior is now unambiguously required, while keeping the spec abstract: describe outcomes and essential design decisions, not code structure. Do not over-specify to a single implementation, and do not leak function names, variable names, paths, or line numbers unless they are undiscoverable external contracts.
4. Preserve everything in the original specification that was already correct; only strengthen what led to the failure.

## Output requirements

- Output the COMPLETE rewritten specification (S'), self-contained and ready for a fresh regeneration agent.
- Keep the same overall structure as the original: the three prose sections (Intent and Rationale, Essential Design Decisions, Undiscoverable Facts) plus the declarative requirements checklist.
- Do NOT include instructions to add, modify, or write tests, and do NOT add verification conditions about test coverage or tests passing.
- The rewritten spec must allow someone to verify correctness WITHOUT reading the actual code.
\end{lstlisting}%
  \end{promptbox}%
  \medskip

\begin{promptbox}[label={app:prompt-vc-extraction}]{Verifiable-Condition Extraction}%
    {\itshape\small Pulls structured, provenance-tagged verification conditions out of the
   session evidence for later checking against regenerated code.\par}%
    \vspace{2pt}%
    \noindent\textcolor{promptrule}{\rule{\linewidth}{0.4pt}}\par
    \vspace{3pt}%
\begin{lstlisting}[style=promptstyle]
Extract structured verification clues from the source coding-agent evidence bundle.

A verification clue is a unique, checkable YES/NO question about behavior, scope, user corrections, or acceptance criteria that generated code C' should satisfy. Existing graders still inspect the final code diff, so each clue must remain concise and independently answerable.

Return JSON with `verification_clue_records`. Each record must include:
- `clue_id`: stable raw id such as "raw_0001".
- `clue`: the verification clue string, phrased as a YES/NO question.
- `provenance`: one or more evidence objects.
- `unsupported_by`: include "final_unified_diff" when the final diff does not directly support the clue.
- `extraction_stage`: use "raw_extraction".
- `pre_dedup_clue_ids`: include the record's raw id.
- `abstraction_level`: one of "user_goal", "behavioral_contract", "api_surface", or "implementation_detail". Classify how specific the clue is. Prefer "user_goal" or "behavioral_contract". Only use "implementation_detail" when the user explicitly requested that exact detail.

Allowed provenance `source_type` values:
- conversation
- metadata
- final_unified_diff
- test_commands
- unknown

Rules:
- Return only clues grounded in the evidence bundle.
- Prefer 5-15 concise clues when available.
- Deduplicate duplicate or near-duplicate clues.
- Phrase each clue as a YES/NO question a reviewer can answer from the code diff.
- Do not include vague clues such as "Does the code work correctly?".
- Prefer outcome-level product or API contract clues over implementation mechanics.
- Include implementation details only when the conversation makes them acceptance criteria.
- Avoid clues about imports, private attributes, helper function names, file layout, or build graph changes unless the conversation makes that exact detail part of the requested contract.
- Apply the counterfactual test: a good clue should be answerable YES for any correct implementation of the user's request, not only the specific implementation in this diff. If a clue would be answerable YES only with the exact same variable names, file layout, or internal API calls, it is too specific — rewrite it as a behavioral or contract-level question.
- Set `not_supported_by_final_diff` whenever no `final_unified_diff` provenance directly supports the clue.

Better clue style:
- Good: "Can personal and project coding-agent memory be routed to the memory service independently behind separate rollout gates?"
- Good: "Does project memory remain project-scoped when routed through the memory service, while personal/session memory remains viewer-scoped?"
- Good: "Do memory-service read failures fall back to directory listing before surfacing a not-found error?"
- Too implementation-specific: "Does AgentMemoryExtension import MemoryClientFactory in a TYPE_CHECKING block?"
- Too implementation-specific: "Is a private `_memory_storage` attribute initialized to None?"
- Too test-specific: "Are two tests named `test_storage_rename_personal_delegates_to_memory` added?"

Exclusions — do NOT include clues about:
- Intermediate agent actions, tool calls, or debugging steps (e.g. "the agent searched for X", "validation was run").
- External system state not part of the diff (e.g. feature-flag or config-service metadata, spreadsheet contents, dashboard outputs, database state, API responses).
- Test execution outcomes (e.g. "all tests should pass", "code should pass linting", "N tests pass"). Test running is handled by a separate grader.
- Test existence or test coverage (e.g. "a test should be added", "tests should be updated"). Test verification is handled by a separate grader.
- Analysis outputs, numeric results, or data artifacts produced during the conversation that are not part of the code change.

Focus exclusively on what the code diff should contain: behavioral changes, structural modifications, API usage patterns, removed/added logic, and correctness invariants.

Conversation context:
```
{conversation_context}
```

Metadata JSON:
```json
{metadata_json}
```

Final unified diff:
```diff
{unified_diff}
```
\end{lstlisting}%
  \end{promptbox}%
  \medskip

\begin{promptbox}[label={app:prompt-test-manifest}]{Test-Manifest Extraction}%
    {\itshape\small Distills a runnable test manifest (shell commands) from the session so
   regenerated code can be checked by execution.\par}%
    \vspace{2pt}%
    \noindent\textcolor{promptrule}{\rule{\linewidth}{0.4pt}}\par
    \vspace{3pt}%
\begin{lstlisting}[style=promptstyle]
Extract a runnable test manifest from source coding-agent validation evidence.

Rules:
- Return only JSON objects with command and category.
- Commands must be self-contained and runnable from the repository root.
- If a command must run from a subdirectory, include an explicit relative `cd path && ...`.
- Prefer observed test tool calls and shell commands over prose summaries.
- Convert observed `run_test`, `run_tests`, and `run_test_with_coverage` tool args into shell commands.
- Read the conversation chronologically. If older validation commands conflict with
  later validation commands, keep only the latest final validation commands.
- Exclude tests for files, classes, or commands that the later conversation says
  were deleted, renamed, replaced, removed, obsolete, or no longer referenced.
- Prefer commands from final "Verification", "Validated", "Done", or summary
  sections over commands from early exploration or intermediate stack edits.
- Prefer concise smoke coverage over exhaustive historical retries. Keep the
  strongest 1-3 final commands unless more are explicitly required.
- Do not output raw JSON tool args as commands.
- Include build/lint commands when they were used as correctness checks.
- Use category "regression" for commands from the original validation flow.
- Use category "new" only for additional validation commands clearly requested or proposed.
- Do not invent build targets, file paths, package scripts, or setup commands.
- Do not convert table names, product abbreviations, or package labels into repo paths.
- Do not include dependency installation, server startup, file inspection, or data-query commands.
- If the conversation says tests passed but does not show the command, return no command for that claim.

Test command conversion examples:
- Observed `{{"test_file_path":"/path/to/repo/frontend/lib/foo/__tests__/FooTest.js"}}`
  -> `jest frontend/lib/foo/__tests__/FooTest.js`
- Observed `{{"test_class":"FooTest"}}`
  -> `test-runner FooTest`
- Observed `{{"test_name":"FooTest::testSomething"}}`
  -> `test-runner FooTest::testSomething`
- Observed `{{"test_file_path":"/path/to/repo/backend/example/tests/test_foo.py"}}`
  -> `python -m pytest backend/example/tests/test_foo.py`
- Observed `{{"file_path":"/path/to/repo/backend/example/tests/test_foo.py"}}`
  -> `python -m pytest backend/example/tests/test_foo.py`
- Observed `{{"test_file_path":"/path/to/repo/backend/example/FooTest.cpp"}}`
  -> no command unless a build target or concrete command is also present.
- Observed a build-system test command with an explicit target (e.g. `build-tool test //project/tests:test_manifest_extractor`)
  -> keep the build-system test command.
- Observed a build-system build command with an explicit target (e.g. `build-tool build //project:run_optimizer`)
  -> keep the build-system build command.
- Observed `cargo test` without a working directory
  -> keep it only if the surrounding context makes the repository-root command correct.
- Observed `run_test` / `run_tests` with empty args `{{}}`
  -> no command.
- Older observed `{{"test_file_path":"/path/to/repo/frontend/lib/foo/__tests__/OldTest.js"}}`,
  but the later conversation says `OldTest.js` was deleted and final
  verification ran `NewTest`
  -> omit `OldTest.js` and keep `test-runner NewTest`.

Observed test commands/tool args:
```
{observed_test_commands}
```

Conversation:
```
{conversation_context}
```
\end{lstlisting}%
  \end{promptbox}%
  \medskip

\begin{promptbox}[label={app:prompt-runtime}]{Runtime Instruction Wrapper}%
    {\itshape\small The instruction wrapper an implementer agent receives at runtime to
   regenerate code from a specification and satisfy its verification
   conditions.\par}%
    \vspace{2pt}%
    \noindent\textcolor{promptrule}{\rule{\linewidth}{0.4pt}}\par
    \vspace{3pt}%
\begin{lstlisting}[style=promptstyle]
You are given a **Reimplementation Spec** — a reimplementation specification plus a verification checklist that describes what changes need to be made to this codebase. Your task is to implement code changes that satisfy the specification and ALL of its verification conditions.

## Reimplementation Spec

{spec}

## Instructions

1. Read the spec carefully - focus on the **Intent and Rationale**, **Essential Design Decisions**, **Undiscoverable Facts**, and the **Verification Conditions** checklist
2. Explore the codebase to understand the current state
3. Implement code changes that satisfy the spec and answer YES to every verification condition
4. The spec describes WHAT should change, not HOW to implement it - use your judgment for implementation details
5. Do NOT over-engineer - make the minimal changes needed to satisfy the spec and its verification conditions
\end{lstlisting}%
  \end{promptbox}%
  \medskip

\subsection{Multi-Tier Verification Graders}
\label{app:prompts-graders}

Three graders evaluate regenerated code $C'$ against the reference change
$C$ and combine into a single weighted score, $\mathit{solve\_score} =
3\cdot\mathrm{flex} + 2\cdot\mathrm{vc} + 1\cdot\mathrm{align}$
(maximum $6.0$). Each continuous grader emits its grade as a final
\texttt{FINAL\_GRADE=<ratio>} line; the ground-truth grader is binary.
Tier labels below follow the grader source files, which number the
ground-truth tier first---the reverse of the paper's ordering.

\begin{promptbox}[label={app:prompt-flex-grader}]{Flexible Test-Execution Grader \textnormal{(Tier 3)}}%
    {\itshape\small Runs the extracted test manifest against $C'$ and reports a continuous
   pass ratio in $[0,1]$.\par}%
    \vspace{2pt}%
    \noindent\textcolor{promptrule}{\rule{\linewidth}{0.4pt}}\par
    \vspace{3pt}%
\begin{lstlisting}[style=promptstyle]
You are an execution verification grader. You run a provided test manifest against generated code C' and may salvage tests only when failures are structural mismatches. You report a continuous PASS RATIO in [0.0, 1.0] (the fraction of runnable tests that pass), not a binary pass/fail verdict. This ratio feeds the solve_score aggregator.

Test Manifest (JSON array): {test_manifest}

Max Salvage Attempts: {max_salvage_attempts}

Diff Number: {diff_number}

Repository Type: {repository_type}
Repository Path: {repository_path}

Instructions:
- Parse test_manifest as a JSON array of objects with command and category fields. category is "new" or "regression".
- If the manifest is empty or invalid, call set_float_grading_result_tool with pass=false and grade=0.0 and the reason "no runnable tests".
- Run each command exactly as provided from the repository root, or from repository_path for git repositories.
- If a test fails because of import paths, symbol names, module names, or constructor/function signatures, you may adapt the test harness and rerun.
- You may adapt imports, module names, function/class names, and signatures only.
- You must not adapt assertions, expected values, or behavioral checks.
- Never modify implementation code.
- If a test command references a test method or class that does not exist, inspect the original/reference patch C by running the version control system's diff command for the reference change {diff_number} via execute_command.
- Mark a missing test as SKIPPED only when the original/reference patch C removed or renamed that test, or removed the feature/behavior that the test covered. Use the reason "test intentionally removed by reference patch". Do not count it as a failure.
- If the reference patch does not remove or rename the missing test/covered behavior, or if the evidence is inconclusive, mark the command as FAIL. Do not skip solely because the generated checkout lacks the test.
- Stop after max_salvage_attempts total adaptations.
- Compute passed_ratio = passing tests / (total tests - skipped tests), the fraction of runnable tests that pass after allowed salvage, a float in [0.0, 1.0]. Report the exact ratio (e.g. 0.5 for 1 of 2) — do NOT round to 0 or 1. If every test is SKIPPED so there are no runnable tests, set passed_ratio=0.0.
- REQUIRED FINAL STEP: call set_float_grading_result_tool exactly once with grade=passed_ratio (the numeric ratio) and pass=true only if all regression tests pass (excluding skipped) and every runnable new test passes after allowed salvage, otherwise pass=false. The grade carries the continuous signal consumed by the solve_score aggregator; the pass flag is only a coarse indicator and is NOT used in the score.
- Wait for the set_float_grading_result_tool response before writing your final verdict.
- REQUIRED: the VERY LAST line of your exit report must be a single machine-readable marker in EXACTLY this format, with no other text on the line: `FINAL_GRADE=<passed_ratio>` where <passed_ratio> is the numeric ratio as a decimal in [0.0, 1.0] (e.g. `FINAL_GRADE=0.5`). Emit it on every path, including when there are no runnable tests (use `FINAL_GRADE=0.0`). This line is parsed by the optimizer and must always be present and exact.
- Exit with the report below.

<format>
  # Flex Test Verdict

  ## Test Results
  | Command | Category | Result | Adaptations | Evidence |
  |---------|----------|--------|-------------|----------|
  | [cmd] | [new/regression] | [PASS/FAIL/SKIPPED] | [none or summary] | [log excerpt or reason for skip] |

  ## Summary
  total: [N]
  passed: [N]
  skipped: [N]
  passed_ratio: [0.0-1.0] (passing / (total - skipped))  -- this value is reported as grade
  new_percentage: [0.0-1.0 or N/A]
  regression_percentage: [0.0-1.0 or N/A]

  ## Result
  passed_ratio=[0.0-1.0] -- [Pass / Fail]
</format>
\end{lstlisting}%
  \end{promptbox}%
  \medskip

\begin{promptbox}[label={app:prompt-vc-grader}]{Verifiable-Condition Check Grader \textnormal{(Tier 2)}}%
    {\itshape\small Checks how many of the extracted verification conditions $C'$ satisfies
   and reports a continuous ratio in $[0,1]$.\par}%
    \vspace{2pt}%
    \noindent\textcolor{promptrule}{\rule{\linewidth}{0.4pt}}\par
    \vspace{3pt}%
\begin{lstlisting}[style=promptstyle]
You are an intent-level verification grader. You check whether generated code C' satisfies concrete verification clues extracted from the source conversation. You report a continuous satisfaction RATIO in [0.0, 1.0] (the fraction of evaluable clues that C' satisfies), not a binary pass/fail verdict. This ratio feeds the solve_score aggregator.

Verification Clues (JSON array of unique clue strings): {verification_clues}

Pass Threshold: {pass_threshold}

Repository Type: {repository_type}
Repository Path: {repository_path}

Instructions:
- Parse verification_clues as a JSON array of unique clue strings. If it is invalid, call set_float_grading_result_tool with pass=false and grade=0.0 and a clear reason.
- If verification_clues is an empty array, do not inspect the diff. Record a SKIP by calling set_float_grading_result_tool with pass=true and grade=0.0 and summary "SKIP: no verification clues to evaluate", then exit with a report that has passed=0, failed=0, skipped=0, total=0, satisfied_ratio=0.0 (N/A), and Result=Skip.
- Use check_changes_since_last_commit to inspect generated changes C'.
- For each clue, read the relevant source files and decide PASS or FAIL.
- Cite file:line evidence when possible; use "not found" evidence for failures.
- IMPORTANT: Your scope is the behavioral correctness of the code diff. Do NOT evaluate:
  - Test existence or test coverage — whether tests were added or updated is handled by the test-execution grader.
  - Test execution outcomes — whether tests pass is handled by the test-execution grader.
  - External system state — feature-flag config, config-service metadata, spreadsheet data, dashboard data.
  If a clue falls into these categories, mark it SKIP and exclude it from the satisfied ratio.
- Compute satisfied_ratio = passed clues / (total clues - skipped clues). This is the fraction of evaluable verification clues satisfied by C', a float in [0.0, 1.0]. Report the exact ratio (e.g. 0.6667 for 2 of 3) — do NOT round to 0 or 1.
- If every clue is SKIP and there are no evaluable clues, record a SKIP by calling set_float_grading_result_tool with pass=true and grade=0.0 and summary "SKIP: no evaluable verification clues", then exit with Result=Skip. Do not mark all-skipped clue sets as Fail.
- REQUIRED FINAL STEP: call set_float_grading_result_tool exactly once with grade=satisfied_ratio (the numeric ratio) and pass=true when satisfied_ratio is at least pass_threshold, otherwise pass=false. The grade carries the continuous signal consumed by the solve_score aggregator; the pass flag is only a coarse threshold indicator and is NOT used in the score.
- Wait for the set_float_grading_result_tool response before writing your final verdict.
- Do not finish, exit, or claim the grading result was recorded until set_float_grading_result_tool has returned successfully.
- REQUIRED: the VERY LAST line of your exit report must be a single machine-readable marker in EXACTLY this format, with no other text on the line: `FINAL_GRADE=<satisfied_ratio>` where <satisfied_ratio> is the numeric ratio as a decimal in [0.0, 1.0] (e.g. `FINAL_GRADE=0.6667`). Emit it on every path, including SKIP (use `FINAL_GRADE=0.0`). This line is parsed by the optimizer and must always be present and exact.
- Exit with the JSON-like report below.

<format>
  # Verification Clues Verdict

  ## Clue Results
  | Clue | Result | Evidence |
  |------|--------|----------|
  | [full clue] | [PASS/FAIL/SKIP] | [file:line or not found or reason for skip] |

  ## Overall
  passed: [N]
  failed: [N]
  skipped: [N]
  total: [N]
  satisfied_ratio: [0.0-1.0] (passed / (total - skipped))  -- this value is reported as grade

  ## Result
  satisfied_ratio=[0.0-1.0] -- pass_threshold=[value] -- [Pass / Fail]

  FINAL_GRADE=[0.0-1.0]
</format>
\end{lstlisting}%
  \end{promptbox}%
  \medskip

\begin{promptbox}[label={app:prompt-gt-grader}]{Ground-Truth Alignment Grader \textnormal{(Tier 1)}}%
    {\itshape\small Issues a binary Pass/Fail behavioral-equivalence verdict comparing $C'$
   against the reference change $C$.\par}%
    \vspace{2pt}%
    \noindent\textcolor{promptrule}{\rule{\linewidth}{0.4pt}}\par
    \vspace{3pt}%
\begin{lstlisting}[style=promptstyle]
You are an expert code verifier. You verify whether generated code C' is behaviorally equivalent to the reference change C for the stated specification, modulo the source conversation context when it is provided.

Task Description: {task_description}

Specification: {spec_description}

Conversation Context: {conversation_context}

Repository Type: {repository_type}
Repository Path: {repository_path}
Reference Patch Available: {has_reference_patch}

Instructions:
- Read the task description, optional specification description, and optional conversation context.
- Use check_changes_since_last_commit to inspect generated changes C'.
- If a reference patch is available, use get_reference_patch to inspect C.
- Read relevant source files before making behavioral claims.
- Trace concrete scenarios and actively look for evidence against your conclusion.
- REQUIRED FINAL STEP (do this exactly once, and BEFORE you call exit): call set_grading_result_tool with pass=true when C' is behaviorally equivalent to C (your verdict is Pass) or pass=false otherwise (Fail). The boolean you pass MUST match the "## Result" line of your verdict below.
- Do NOT call exit until set_grading_result_tool has been called. If you have already written your verdict text, still call set_grading_result_tool first — a verdict that is not recorded through set_grading_result_tool is counted as a Fail regardless of your written conclusion, so you must always emit it. Emit it even when you are highly confident or when the answer seems obvious.
- Only after set_grading_result_tool succeeds, call exit and include the structured reasoning trace below (its "## Result" must agree with the pass value you sent).

<format>
  # Verification Verdict

  ## Issue/Change Summary
  [2-3 sentence summary]

  ## Function Trace Table
  | Function/Method | File:Line | Behavior (VERIFIED) |
  |-----------------|-----------|---------------------|
  | [function]      | [file:N]  | [actual behavior]   |

  ## Behavioral Comparison
  [Concrete scenarios comparing C and C']

  ## Conversation Alignment
  [How the verdict accounts for provided conversation context, or "No conversation context provided"]

  ## Alternative Hypothesis Check
  [Evidence searched for and found]

  ## Critical Differences
  [Behavioral issues or "None identified"]

  ## Result
  [Pass / Fail] -- Confidence: [HIGH / MEDIUM / LOW]
</format>
\end{lstlisting}%
  \end{promptbox}%
  \medskip

\end{document}